# Correlation between tides and seismicity in Northwestern South America: the case of Colombia

*Gloria A. Moncayo[a], Jorge I. Zuluaga[a] and Gaspar Monsalve[b]*

[a] Solar, Earth and Planetary Physics Group, Computational Physics and Astrophysics Group, Instituto de Física-FCEN, Universidad de Antioquia, Calle 70 No. 52-21, Medellin, Colombia

[b] Departamento de Geociencias y Medioambiente, Facultad de Minas, Universidad Nacional de Colombia, Carrera 80 No. 65-223, Núcleo Robledo, Medellín, Colombia

**Abstract**

We present the first systematic exploration of earth tides-seismicity correlation in northwestern South America, with a special emphasis in Colombia. For this purpose, we use a dataset of ~167,000 earthquakes, gathered by the Colombian Seismological Network between 1993 and 2017. Most of the events are intermediate-depth earthquakes from the Bucaramanga seismic nest and the Cauca seismic cluster. For this purpose, we implemented a novel approach for the calculation of tidal phases that considers the relative positions of the Earth-Moon-Sun system at the time of the events. After applying the standard Schuster test to the whole dataset and to several earthquake samples (classified by time, location, magnitude and depth), we found strong correlation anomalies with the diurnal and monthly components of the tide (global $\log(p)$ values around -7.0 for the diurnal constituent and -12.1 for the monthly constituent), especially for the intermediate depth events. These anomalies suggest that around 16% of the deep earthquakes in Colombia may be triggered by tides, especially when the monthly phase is between 350º-10º. We attribute our positive results, which favor the tidal-triggering hypothesis, in contrast to previous negative ones to: 1) the size of our dataset, and 2) the method we used to calculate tidal phases. Anyone willing to reproduce our results or to apply our methodology to custom datasets can use the public information system **tQuakes** that we developed for this work.

# 1. Introduction

The idea that tides can trigger earthquakes (among other geophysical phenomena) is not new (Perrey, 1855, Knott, 1896, Schuster, 1897) and in recent years it has been widely investigated by geophysicists worldwide. However, proving a causal connection between tides and seismicity, either by statistical analysis of seismic events or modelling the physical mechanisms at play, has been elusive and somewhat contradictory.

In some cases, the correlation seems to be weak, or at least not as significant as expected (Heaton, 1982, Hartzell and Heaton, 1989; Vidale et al., 1998; Beeler and Lockner, 2003; Fischer et al., 2006; Ader and Avouac, 2013). In others, the use of larger and more precise databases, has led scientists to stronger cases in favor of the tidal-triggering hypothesis (Wilcock 2001, 2009; Tanaka et al., 2002, 2004, 2006; Tanaka, 2010; Cochran et al., 2004; Cadichenau et al., 2007; Métivier et al., 2009; Kolvankar et al., 2010; Chen et al., 2012; Vergos et al., 2015; Xie et al., 2015; Arabelos et al., 2016; Ide et al., 2016), especially under very specific circumstances. Thus, for instance, earthquakes can be tidally triggered if the small stress involved at tides, acts in the same direction as tectonic stress (Tanaka et al. 2002, 2004, 2006; Tanaka, 2010, 2012). Additionally, tidal triggering may also occur in regions subjected to a critical state of stress (e.g. Tanaka et al., 2002, 2004, 2006; Tanaka, 2010, 2012; Arabelos et al., 2016) or in regions around large events and/or in previous years to their occurrence (Tanaka et al. 2006; Ide et al., 2016).

The degree of correlation between earthquakes and tides reported in literature depends on the so-called *constituent of the tidal oscillations*, namely the Fourier component of the tidal signal having precise periods (semidiurnal, diurnal, fortnightly, monthly). Thus, for instance, Beeler and Lockner (2003) reported a weak correlation of tides-earthquakes with the diurnal component. Others confirmed this result and extended it to the semidiurnal constituent (Cochran and Vidale, 2007; Métivier et al., 2009; Chen et al., 2012) but discarded any observable correlation with the fortnightly (biweekly) constituent.

The west coast of South America belongs to the circum Pacific belt and has a significant seismic activity. In this region subduction is segmented (Barazangi and Isacks, 1976) and their associated tectonic stresses point mainly in the direction of maximum tidal stress (E-W direction). In Colombia, for instance, the subducting Nazca Plate seems to be segmented in a flat slab to the north and a steeper slab to the south (Vargas and Mann, 2013; Chiarabba et al., 2015; Syracuse et al., 2016), configuring a region with frequent and intense seismic activity. These conditions point out to the fact that northwestern South America could offer an interesting opportunity to study the correlation between tides and seismic activity in a diverse and complex tectonic environment. However, with the only exception of the work by Gallego et al. (2013) no systematic analysis has yet been performed in Central and South America.

This work is different from the already abundant literature in the field, in four main aspects: 1) we explore a region with a diverse tectonic setting that has been so far overlooked and that could hide interesting aspects of the tides-seismicity correlation (see Section 2); 2) we introduce novel techniques to compute the phases of the main tidal components, with special emphasis in their relationship with the astronomical configuration of the Earth, the Moon and the Sun (Section 3);

3) we explore in depth the multidimensional space (geographical location, magnitude, depth and time) where seismic events and tides occur (Section 4); and last but not least, 4) we introduce the first public tool, the `tQuakes` information system, that could help other researchers in the field to explore the tides-seismicity correlation with customized data sets (Section 5). Our aim here is not to demonstrate a definitive correlation between tides and seismicity in northwestern South America, but to reveal some hints of such a connection that may motivate the development of other in-depth studies in the field in this region.

## 2. Tectonic setting

The northwestern corner of South America is an active and intriguing area in terms of tectonic activity. Figure 1 shows the main structural features and the tectonic setting of the area. The Nazca plate subducts in a roughly W-E direction, at a rate of ~5.3 cm/yr relative to stable South America (Sella et al., 2002). In the northern coast of Colombia, the Caribbean plate seems to subduct at a shallow angle and an average rate of 1 - 2 cm/yr. (Taboada et al., 1998; Trenkamp et al., 2002).

Besides the main tectonic plates (Nazca, Caribbean and South American) that interact in the study region, two tectonic blocks should additionally be considered: The Panama-Choco Block and the North Andean Block (NAB) (See Cortés and Angelier, 2005 and references there in). This tectonic setting has resulted in the accretion of several exotic terrains (Taboada et al, 1998, 2000, Cediel et al., 2003) and the formation of three mountain ranges (Western, Central and Eastern Cordilleras). The main active fault systems in Colombia are mostly parallel to the cordilleras (see **Figure 1**); they include the Llanos Foothills System, the Magdalena Valley Fault System (MVF) and the Romeral Fault System (Taboada et al, 1998, 2000).

Several studies suggest the existence of two different subduction segments separated by an E-W slab tear at a latitude of about 5ºN, with a steep segment to the south and a flat segment to the north (Pennington, 1981; Vargas and Mann 2013; Yarce et al., 2014; Chiarabba et al., 2015; Syracuse et al., 2016). Associated to these segments there are two remarkable seismic clusters: the Bucaramanga nest and the Cauca cluster. In Supplementary **Figure S1** we show the spatial distribution of earthquakes related with them. The Bucaramanga seismic nest configures one of the most seismically active regions in the world. The events in this nest, whose hypocenters have an average depth of ~140-160 km (Taboada et al., 1998; Prieto et al., 2012), are related to the subduction of either the Caribbean and/or the Nazca plates (Pennington 1981; Taboada et al., 2000; Zarifi and Havskov, 2003; Cortes and Angelier, 2005; Zarifi et al., 2007; Prieto et al., 2012). The Bucaramanga nest occupies the smallest volume of any intermediate-depth seismic nest in the world (Taboada et al., 2000; Zarifi and Havskov, 2003; Prieto et al. 2012). About 400 km southwest of the Bucaramanga nest there is another source of intermediate-depth events: the Cauca cluster (Vargas and Mann, 2013;

Chang et al., 2017), which seems to be related to the subduction of the Nazca Plate (Tabares Ocampo et al., 1999; Taboada et al., 2000; Cortes and Angelier, 2005; Chiarabba et al., 2015). Recent studies have suggested that the two segments on which the seismic events of the Bucaramanga nest and the Cauca cluster arise, are disconnected below 50-100 km depth (Syracuse et al., 2016), due to the existence of a slab tear (Vargas and Mann, 2013; Syracuse et al., 2016).

## 3. Materials and Methods

### 3.1. Dataset

Our dataset contains 167,162 earthquakes, covering 24 years between June 6, 1993 and August 14, 2017. The earthquakes have been recorded by the Colombian National Seismological Network (RSNC for its acronym in Spanish), a network of countrywide detectors operated by the Colombian Geological Survey.

The earthquake description, including origin time in UTC, epicenter location and depth, local magnitude, moment magnitude, location errors, among other properties, was obtained from the RSNC website and stored in a MySQL database. The geographical location of the events ranges in longitude from -90º to -66º and in latitude from –7º to 15º (see **Figure 2**). The complete list of the earthquakes used in this work, along with their geophysical and related tidal properties, can be browsed in the **tQuakes** information system, especially developed for this work, and now publicly available in http://seap-udea.org/tQuakes (see Section 6).

The RSNC reports both local and moment magnitudes for only 4,604 earthquakes. We will only consider here the local magnitude unless stated otherwise. Although moment magnitudes are the most robust estimate of the earthquake energy (Ristau et al., 2003), for studying the correlations with tides the most critical properties of the events will be location and time. Magnitudes are used here just to classify and filter the events. Thus, for instance, when a magnitude threshold is applied to the database, we use local (which is always available) instead of moment magnitude. We have verified that less than 10% of the earthquakes have moment magnitudes significantly larger than the local ones. With samples containing hundreds to thousands of earthquakes, the statistical effect of including or discarding events with large moment magnitude is small and will not affect our conclusions.

### 3.2. Cluster analysis

It is well known that after a large earthquake, aftershock sequences for a restricted interval of time and space may occur (Utsu, 1969). To avoid any statistical bias introduced by aftershocks (clustering biases), we need to remove them from the database. Although some authors argue that removing aftershocks is not necessary for studying the shortest constituents of the tidal signal, namely the diurnal and semidiurnal component (Jeffreys 1938; Métivier et al., 2009), since we intend to analyze correlations with long period components, the declustering of the sample is mandatory.

Distinguishing foreshocks, mainshocks and aftershocks is challenging (their very definitions are mostly subjective, Talbi et al., 2013). The most popular criterion to identify aftershocks is their proximity in time and space (Christophersen et al., 2011). Several algorithms for declustering, using this spatio-temporal criterion, have been widely described and tested in literature (see e.g. van Stiphout et al., 2012; Talbi et al., 2013 and references there in). In our case, and for practical purposes, we will restrict to the well-known Reasenberg algorithm (Reasenberg, 1985; van Stiphout et al., 2012). The algorithm defines an interaction zone around each main earthquake. All the events that happen after de mainshock inside this area are potentially considered aftershocks. If they fulfill several additional criteria (related to their magnitude and times) the algorithm classifies them as members of a cluster. The interaction zone is parameterized with two properties: a characteristic spatial-scale ($R_{fact}$) and a maximum time-scale ($\tau_{max}$, also called the Omori's law parameter). In our case, we have chosen $R_{fact}$=10 km and $\tau_{max}$=10 days, which are standard values used for declustering low-magnitude events (van Stiphout et al., 2012; Talbi et al., 2013). We have tested different values of $R_{fact}$ and $\tau_{max}$ in the ranges found in the literature and verified that the resulting clusters do not change significantly with respect to the reference parameters.

After applying the Reasenberg's algorithm to our unfiltered earthquake database, we identified 925 clusters, containing a total of 58991 events. These events represent 36% of the total sample. Most of them were tagged and excluded from the statistical analysis, leaving only the largest event in each cluster. The earthquakes plotted in **Figure 2** are the 108459 events that constitute our declustered sample.

The seismic network in Colombia has considerably improved during the last 15 years (starting around spring 2003). As a result, the distribution of events in time and magnitude is not entirely homogeneous. In **Figure 3** we present a scatter plot of the local magnitude and time of the earthquakes in our database. We identify two distinct periods. On the one hand, we identify a time interval, which we classified as period A, that contains events between 6 June 1993 and approximately 21 March 2008; a noticeable lack of low magnitude events ($M_l$ < 2) is observed during this period. On the other hand, period B covers the interval between 21 March 2008 and 13 august 2017. The differences in magnitude distribution among these periods are attributable to the expansion of the station network between 2003 and 2008 (Rengifo et al., 2003).

To avoid statistical artifacts introduced by the incompleteness of the period A sample, we will use two strategies: 1) when the whole sample is analyzed, only earthquakes with magnitude $M_l$ >2.0 are considered; and 2) when including events with magnitude $M_l$ < 2.0 we performed separate analysis for both periods.

### 3.2.1. Other causes of clustering

Aftershocks are not the only clustering cause of the dataset. Other effects, such as fluid migration, seismicity crisis, volcanic activity, mining, etc. (see e.g. Simpson et al., 1988), may contribute to group events at specific times and places, introducing numerical artifacts in the tidal phase analysis.

Specifically in Colombia, we can find some possible sources of seismicity that are not tectonic in character. The existence of dams, natural lakes, volcano activity, and also the mining activity, can produce seismicity without a tectonic process. In the area of the Bucaramanga nest, the presence of dams has raised the question of the possibility of induced seismicity; however, up-to-date there is no evidence of an increase in seismicity that could be related to the dam activity. According to Simpson et al., (1988), the seismicity caused by the activity of dams would be limited to shallow events, approximately up to about 30 km. Most earthquakes in the Bucaramanga nest are of intermediate-depth (> 70 km, e.g. Zarifi and Havskov, 2003; Prieto et al., 2012). Although seismic events in dam regions could eventually reach high magnitudes (the best documented case is the Ml=6.5 event in the region of the Koyna dam in India reported in Gupta, 2002; Telesca, 2010), if those events are more common, our ability to identify them in a large database is still very limited.

Several active volcanoes exist in the northern Andes. Volcanoes showing recent activity are continuously monitored by the Colombian Geological Survey through their Volcanological and Seismological Observatories. It can be said that the activity of these volcanoes is mostly restricted to their areas of influence. Still, some volcano-tectonic seismic events of relatively large magnitudes can also be recorded by the National Seismological Network. Thus, some of our events (especially in the Cauca region and in the area of the Colombia – Ecuador border) may also be volcano-tectonic in origin. It is not easy to distinguish these events from a pure tectonic event. Still, a strong volcano-tectonic event may also be tidally-triggered, and their inclusion, even if by accident, may also help in our analysis.

The mining activity or the hydrocarbon exploration can also contribute with some induced seismicity in the region. Events with this origin are characterized by shallow depths and low magnitudes (Simpson et al., 1988; Ruiz-Barajas et al., 2017). Although it is difficult to recognize if these types of induced events are in our large catalogue database, given their low magnitude, probably most of them were not included in our analysis or even they were probably missed by the National Seismological Network.

In the Eastern Llanos Basin, to the east of the Llanos Fault System, (Figure 1) a small cluster of anomalous seismicity (Figure 2), which was not known before 2013 and which is probably associated to hydrocarbon operations, was already identified by Gomez-Alba et al., 2015. However, the region is far away from the largest concentration of seismic events in Colombia. Although we probably do not filtered out these events, because of their small magnitudes they were excluded in most of the correlation analysis we perform below.

### 3.2.2. Testing the declustering algorithm

To test the quality of the aftershock removal algorithm and eventually prevent artifacts introduced by other clustering causes, we have performed several statistical tests on the declustered sample.

It is well known that seismic events should follow a spatially inhomogeneous, temporally homogeneous Poisson process (SITHP). Multiple statistical tests have been devised and widely evaluated in the literature to test if a seismic dataset fulfills this model (for a review see eg. Luen & Stark, 2012 and references there in). Since the aim of this work is not performing a rigorous analysis of the quality of the declustering process, but only avoiding as much as possible the introduction of numerical artifacts when analyzing the tidal phases of the events in the catalogue, we will restrict our efforts to compute two basic statistics on different spatiotemporal windows of the unfiltered and declustered dataset:

1. **Interval counts, $N_i$**. After extracting a sample of events in a region of the parameter space (time, magnitude and depth interval, and geographical region), we divided time in *K* equal subintervals and count the number of events falling into them ($N_i$). The size of the subintervals is fixed in 1 day (which provides a non-negligible event rate for the studied period). If we restrict to a specific geographic region (e.g. Bucaramanga seismic nest), $N_i$ should follow a multinomial distribution (if the process is homogeneous temporal Poissonian) and the temporal rate $\lambda(t)$ would be constant. In **Figure 4** we show a plot of $N_i$ for different time intervals in the Bucaramanga seismic nest (the origin of most of the events in our catalogue). We have restricted these subsamples to events with magnitudes larger than 2 (see completeness analysis below) and depths larger than 20 km.

2. **Kolmogorov-Smirnov statistics of the rescaled time intervals**. For the same subsamples represented in Figure 4, we obtain the time between events and apply the time-rescaling theorem (Brown et al. 2002, Pillow, 2009). For this purpose, we perform the map:

    $$z_i = 1 - e^{-\Lambda_{i-1}(t_i)(t_i - t_{i-1})}$$

    Where $\Lambda_{i-1}(t_i)$ is the difference between the cumulative rate of the process at times $t_{i-1}$ and $t_i$. We estimate $\Lambda_{i-1}(t_i) \approx N_i$. According to the time-rescaling theorem, if the process is Poissonian, $z_i$ should be distributed

uniformly between 0 and 1 (null hypothesis). We test this hypothesis for each sample calculating the Kolmogorov-Smirnov statistics (KS) and the corresponding p-value, $p_{KS}$. We included within the legends of Figure 4, the value of these quantities for the analyzed subsamples.

We classified the possible subsamples as regular and pathological. **Figure 5** shows some regular subsamples, which are characterized by the fact that the unfiltered sets show a significantly larger number of events (as expected), a larger dispersion in the average event rate, and lower p-values. In contrast, the declustered catalogues for the same samples, although having a lower event rate, are more consistent across the years and have larger p-values, which are consistent with the null-hypothesis (time homogeneous Poissonian process).

In contrast, in **Figure 5**, we show examples of pathological subsamples. In one case (left panel in **Figure 5**) the declustering algorithm seems to be unable to remove the foreshocks happening a few days before an $M_L$=4.5 earthquake. As a result, the event rate, two days before the main shock, reaches a peak more than twice higher than the average rate, in both the unfiltered and the declustered cases. On the other hand (right panel in **Figure 5**), we have cases where the declustering algorithm produces a systematic suppression of the event rate, due, apparently, to numerical artifacts.

In summary, our statistical tests show that the declustering algorithm we used for obtaining the final catalogue, is reliable enough for the purposes pursued in this work. However, the algorithm is unable to remove "localized pathologies", such as those presented in **Figure 5**. Specifically, the existence of large spikes in the event rate bias our tidal correlation statistical tests, potentially favoring the hypothesis of tidal triggering (see below). A suppression in event rates, on the other hand, has no positive effect at reinforcing the tidal triggering hypothesis.

To avoid potential biases, introduced in our tidal correlation analysis by those anomalous spikes, we have devised an additional "brute force" filter for our declustered catalogues. Once a declustered catalogue is obtained after applying the Reasenberg algorithm, we calculate the event rate (in earthquakes per day) and estimate its standard deviation ($\sigma_\lambda$) within the time window of the sample. For those days where the observed event rate is larger than $1.5\,\sigma_\lambda$ , we remove all the events from the resulting catalogue.

### 3.3. Magnitude distribution

To assess the completeness of our earthquake database, and in general the regional seismicity, it is customary to compute the cumulative magnitude distribution (see **Figure 6**) and fit it to the Gutenberg-Richter Law (GRL, Gutenberg and Richter, 1944):

$$N\ (>M) = N_o 10^{-bM}$$

Here *N* is the number of events with magnitude larger than *M* and $N_o$ and *b* are free parameters.

We find in our case that the distribution of magnitudes in both, the complete and the declustered databases, follows the GRL with b-values of 0.954 and 0.913 respectively. Previous analysis that were performed on smaller samples, especially for period A (see Figure 3), obtained very different b-values. Thus, for instance, Franco et al. (2003) reported a b-value for Colombian earthquakes of 0.84, while Caneva (2002) reported a b-value of 0.6 for events with depths less than 100 km and for the non-cumulative case, whose value increases once they included events at depths larger than 100 km. Our result not only represents an up-to-date b-value for the region, but it also suggests that the database we are working with is more complete. Our b-value is also more consistent with the type of source environment of the region, where most of the seismic events correspond to those of tectonic character. In our case, the intermediate b-value that we found, would indicate an association to strike slip events (Schorlemmer et al., 2005). Recently, Petruccelli et al. (2018), found that the b-value depends harmonically on the angle of rake, finding large b-values for normal mechanisms and low b-values in the case of reverse ones. Our intermediate b-value is consistent with their results.

As observed in **Figure 6**, both the full database and the declustered sample, are complete for magnitudes larger than and approximately equal to 3.0. This is consistent with the fact that during period A low magnitude events were not detected by the network. This result agrees with previous analysis of the same dataset (Cardona et al, 2005). The peak in the blue curve of the bottom panel of **Figure 6**, arises from the fact that after the declustering procedure, large magnitude earthquakes (that are more prone to aftershocks) are slightly over represented.

It is interesting to notice that most of our earthquakes have magnitudes $M_L$<3.0 (see top panel in **Figure 6**). However, considering that some previous results have suggested that the correlation between tides and seismicity occur especially for low magnitude events (Métivier et al., 2009) of thrust fault type (Tanaka et al., 2002), the fact that our sample is underrepresented at low magnitudes will not contribute to falsely reinforce the hypothesis we want to test. Instead, it suggests that any hint of a correlation would be amplified if we include all the events. Other authors have observed a correlation for large events, $M_L$> 5.0 (e.g. Cochran et al., 2004; Tanaka et al. 2002; 2010), for which our sample is rather complete. In summary, the completeness of our sample will not impact significantly the conclusions or our statistical analysis (see Section 3.6).

### 3.4. Depth distribution

Depths of earthquakes in our database range from 0 to 700 km. Approximately 50% of the events correspond to intermediate-depth earthquakes (depths ~120 km), located mostly around the Bucaramanga nest. Roughly 50% of the earthquakes (~50,000) are shallow or crustal events at depths of 0-70 km with magnitudes $M_L<4.0$. Among them ~36,000 are very shallow events with depths lower than 20 km. Significant errors in earthquake depths are observed among a non-negligible fraction of the database. However, we have verified that for more than half of the events errors are below 10 km. Only ~3% of the events have errors larger than 50 km.

To illustrate the global properties of the magnitude and depth distribution as a function of geographical position, in **Figure 2** we showed contour maps of the median depth and maximum magnitude across the geographical area considered in this work. The Bucaramanga seismic nest and the Cauca cluster are clearly identifiable in the median depth map (top panel of **Figure 2**). Additionally, some deep earthquakes (depth ~140-190) in northern Colombia could be associated to the subduction of the Caribbean plate (Malave and Suarez, 1995). Most of the low-magnitude shallow events happen in the central region of the country, around the cordilleras (Pulido, 2003).

In the case of the Bucaramanga seismic nest, as mentioned by Prieto et al (2012) and references therein, the mechanisms generating intermediate to deep seismicity have not been well understood; however, two main types of mechanism have been proposed: dehydration embrittlement and thermal shear runaway instability. Also, mechanisms like slab detachment, tearing or contortions can be associated with the origin of seismicity in some earthquake nests (Zarifi and Havskov, 2003; Prieto et al., 2012), but probably not for the Cauca cluster, where a continuous seismic zone was defined (Chang et al., 2017). However, Chang et al., 2017 suggest a possible dehydration embrittlement for some shallow earthquakes in the Cauca cluster.

### 3.5. Calculation of tides

To assess the correlation between seismicity and tides, we compute for each event in our sample, the tidal signal at the place (latitude, longitude and depth) and around the date of the earthquake. For this purpose, we use the *Earth tide data processing package* ETERNA 3.40 (Wenzel, 1996, 1997b). The package uses a high-order tidal potential catalogue (Hartmann and Wenzel, 1994, 1995) able to predict synthetic earth tidal signals with an accuracy better than 0.01 nm/s$^2$.

We calculate the time series of different components of the tide (gravitational acceleration, vertical displacement, vertical strain, horizontal strain in the direction of azimuth zero and horizontal strain in the direction of 90 degrees), 90 days before and after the earthquake, with a cadence of 30 minutes. In **Figure 7** we show examples of the resulting time series; a visual inspection of the signal reveals what we will call the "*principal components*" in the tide (tidal waves in Wenzel, 1997a):

(1) the semidiurnal lunar dominant wave (M2), which is responsible for the oscillating nature of the signal.

(2) The diurnal luni-solar wave (K1), which together with the diurnal lunar principal wave (O1) is responsible for the alternating daily maxima in the signal (inset panels).
(3) The fortnightly wave (Mf), which is responsible for the ~13-day modulation of the signal (inset panels).
(4) The monthly lunar (Mm), which is responsible for the ~27-day alternating maxima in the modulation signal (main plots).

Naturally, a much larger number of components have been used to compute the synthetic signal. However, the highlighted wave components produce the most noticeable differences in tide intensity, which ultimately will be responsible for the potential triggering of earthquakes. Once the time-series of the tide was computed for a given earthquake, we proceed to compute the "phase" of the principal components at which the earthquake happens.

The calculation of the tidal phase has varied in the literature since the first attempts to look for a correlation between tides and seismic activity. By definition, the tidal phase of a constituent wave (semidiurnal, diurnal, fortnightly, monthly, etc.) is a measure of the state of the strain and stress caused by the corresponding wave. Thus, for instance, a tidal phase of 0 (0.5) for the monthly component corresponds to a time when the monthly component reaches a maximum (minimum) in the cycle. But computing the tidal phase of the components from the signal time-series in such a way that it reflects the total strain, is not an easy task.

Most authors (Heaton, 1975, 1982; Tsuruoka et al., 1995; Jentzsch et al., 2001; Cochran et al., 2004; Cochran and Vidale, 2007; Tanaka, 2010, 2012; Tanaka et al 2002, 2006) calculate the phase using a linear interpolation between 0º and 180º between the latest maxima and minima around the earthquake occurrence, respectively (we use here a similar definition). Metivier et al (2009) performed a Fourier decomposition of the tidal signal and identified the phase of each wave corresponding to that of the Fourier components. A simpler method was devised and used more recently in Vergos et al., (2015) and Arabelos et al., (2016) in which the phase is calculated as the fraction of the corresponding wave period elapsed since a reference time, irrespective of the amplitude of the tidal signal. Other authors (Wilcock, 2009; Stroup et al., 2009) have considered more carefully the role of the value of the signal to assign the phases to the earthquake. Interestingly, with the only exception of the works by Kolvankar et al. (2010) and Iwata and Katao (2006), which consider the moon phase and its correlation with the origin time of earthquakes, most, if not all methods used in literature, calculate the phase without considering the specific configuration of the Earth-Moon system. This is especially important in the case of the diurnal and monthly components. Naturally, the tidal signal contains some information about this configuration, but not always the relative position of the bodies is correlated with the signal intensity (see below). Moreover, we claim here that using the astronomical configuration to compute the tidal phases is not only desirable, but probably the proper way to do it.

Our novel approach for the calculation of the tidal phases consists on computing the phase of each wave at the time of the earthquakes, consistently with the state of the tidal stresses at the place and time of the event and the astronomical configuration around the corresponding date. A graphical illustration of the method and the resulting phases for the case of the 1999 Quindío event is shown in **Figure 8**.

In a first step, we compute the position of the maxima (minima) of the time series (red dots in **Figure 8**, both in the inset panel and in the main plot). We call these the "semidiurnal maxima (minima)". In a second step, we use the position of maxima (minima) to draw what we called the "carrier wave" of the semidiurnal oscillation (blue continuous line in the inset panel of **Figure 8**). For that purpose, we perform a softening of the position of the semidiurnal maxima (minima)[1], creating a curve that passes between them. This wave has its own maxima and minima that we identify in a third and final step (cyan squares and green triangles in the main plot of **Figure 8**).

Once the maxima and minima of the signal and those of the corresponding carrier wave are identified, we proceed to calculate the phase of each principal wave following these steps:

(1) **Semidiurnal phase**: the phase of this wave is computed as the ratio of the time elapsed from the preceding semidiurnal maxima (red dots in the inset panel of **Figure 8**) to the time between the two successive semidiurnal maxima enclosing the origin time of the event (pink shaded region in the inset panel of **Figure 8**).

(2) **Diurnal phase**: the phase of the diurnal wave is computed by searching the signal maxima closest to the time of culmination of the Moon (time when the moon is highest in the sky at the earthquake epicenter). We call the maxima fulfilling this condition "tidal diurnal maxima". The phase is computed again as the ratio of the time elapsed since the latest diurnal maxima and the origin time of the earthquake, to the time between successive diurnal maxima.

(3), (4) **Fortnightly and monthly phases**: the monthly phase is computed with respect to the highest tides in the month (maxima in the carrier wave). These tides usually occur within a few days of the perigee (solid vertical gray lines in **Figure 8**). For that purpose, we first identify the closest maxima of the carrier wave to the first perigee in the observing window (90 days before the event). From that time, all the alternating maxima (green squares in **Figure 8**) are identified as the starting time of what we call the "*tidal anomalistic month*" (time when the moon is close to the perigee and the tides are maximum). The ratio of the time elapsed since the closest previous highest peak (green square) to the time of the earthquake and the time between the consecutive peaks enclosing the event (dash filled band in

---

[1] The softening of maxima positions is performed using a Butterworth filter of 8$^{th}$ degree with cutoff frequency of 0.125 days$^{-1}$.

**Figure 8**) defines the monthly phase. The fortnightly phase is computed in a similar fashion using the consecutive maxima of the carrier wave.

The 1999 Quindío event happened when the vertical displacement at the earthquake hypocenter was at a maximum (semidiurnal phase near 0) and around 12 hours before the diurnal maximum, namely diurnal phase ~0.5. Normally, phases are measured as an angle. In this work, we will report phases as non-dimensional quantities in the interval [0,1], where 0 would correspond to 0 degrees and 1 to 360 degrees. The event occurred at a time in the month when maximum tides where close to a minimum with respect to previous weeks (fortnightly phase ~0.5) and almost 10 days after the beginning of the tidal anomalistic month (monthly phase ~0.3).

The main advantage of our approach for calculating the tidal phases, with respect to those used in previous works, is that it considers the astronomical setting on which the earthquakes happened, so the phase is identified in a much clearer connection with the position of the moon relative to the geocenter (perigee) and the event location on Earth (time of culmination). In doing so, the probability that the phase reflects the state of the actual tidal stresses, is much larger.

As an example, we show in **Figure 9** a histogram of the monthly phases of around 35,000 earthquakes observed in Colombia during the interval studied here. We clearly observe a statistically significant anomaly in the distribution of monthly phases, with 16% more earthquakes happening at phases between 350 and 10 degrees than those happening with phases around 180 degrees. The "significance" of the difference is around 3-4 sigma, which constitutes a strong hint of a correlation between tides and seismicity in the region.

One way to test if our approach is better than others is to verify if the computed phases reflect the stresses around the time of the earthquake. Thus, for instance, for the monthly phase, a small phase should be normally associated to the largest values of the tidal stresses in the month. In **Figure 10** we illustrate our approach by calculating the correlation between the tidal phase and two astronomical properties: we show the value of the monthly phase as a function of the time between the last perigee and the earthquake. In the upper panel, we have used a conventional method to calculate the monthly phase, namely, we calculate the phase with respect to the maximum of the carrier wave closest to the beginning of the synodic month (the maximum closest to the full moon). As shown in this figure, the phase value is not correlated at all with the time after perigee, meaning that earthquakes having a given monthly phase, eg. a phase close to 0 degrees, could happen at any time during the lunar month and hence will correspond to any value of the maximum tidal stress. On the other hand, when the monthly phase is calculated using our methodology, a strong correlation between phase and time after perigee is observed (lower panel in **Figure 10**). In our supplementary **Figure S2** we plot the value of the maximum daily vertical displacement as a function of the tidal monthly phase as calculated with our method, obtaining that the maximum vertical displacement occurs at low and large values of the monthly phase (near 0 and 360 degrees), precisely when the moon is closest to the perigee. This property is a good indication that the value of the monthly phase, as computed with our method, is a good indicator of the level of tidal stresses experienced at the earthquake hypocenter and around the origin time of the event.

### 3.6. Statistical analysis

Testing the correlation between seismicity and tides (a complex cyclic phenomena) is analogous to test if an unbinned signal has any kind of periodicity. To set up a proper statistical test we need two basic conditions: (1) a null hypothesis, namely "seismicity in a large geographical area do not have a periodicity" (at least for the frequencies associated to tides) and (2) a proper statistic to test the hypothesis.

In this work, we use a form of the "Rayleigh' test" (Brazier, 1994; Agnew and Constable, 2008; Mardia, 1972) also known in literature as the "Schuster test" (ST).

ST relies on the calculation of the statistic $D^2$ (Ader and Avouac, 2013):

$$D^2 = \left(\sum_{i=1}^{N} \sin \theta_i\right)^2 + \left(\sum_{i=1}^{N} \cos \theta_i\right)^2$$

where $\theta_i$ is the phase of the principal wave to be tested (see previous section) as calculated for each earthquake in a sample.

In the asymptotic limit ($N \to \infty$) the value of $2D^2/N$ follows a $\chi^2$ distribution with 2-degrees of freedom (Brazier, 1994) and hence the complimentary cumulative probability distribution (ie. the probability that the actual value of $D$ be larger than the sample value $D_s$) is given by:

$$p \stackrel{\text{def}}{=} P(D > D_s) = e^{-D_s^2/N}$$

$$\log p = -\frac{D_s^2}{N};$$

we call this the "p-value" of the ST. If for a given tidal component, the sample value $D_s$ is small, the p-value of the ST will be large. In this case, we cannot discard the null-hypothesis and hence, a correlation between tides and seismicity (at least for the chosen tidal wave) cannot be neither discarded, nor proved. However, if the sample value $D_s$ is large, the p-value will be small and we can say that the probability that the sample be drawn from a purely random distribution (the null-hypothesis) is small to a confidence level equal to *1-p*. We interpret here a small p-value of ST as an indicator of an anomalous periodicity among earthquakes that could be further investigated. We call this effect a "correlation anomaly".

The ST has been widely used in the literature of tides-seismicity correlation (e.g. Heaton, 1975; 1982; Hartzell and Heaton, 1989; Tanaka et al., 2002, 2006, 2010, 2012; Cochran et al., 2004, Cadicheanu et al., 2007; Ader and Avouac, 2013,

Brinkman et al., 2015; Vergos et al., 2015; Arabelos et al., 2016; Schuster, 1897; Heaton 1975, 1982; Tsuruoka et al., 1995; Emter, 1997; Tanaka et al., 2002, 2006, Cadicheanu et al., 2007). The p-value of the ST has proven to be very useful to look for a connection between both phenomena. Thus, for instance, Tanaka et al., (2002, 2004, 2006, 2010, 2012) and Cadicheanu et al. (2007), have shown that the p-value, in specific geographical areas, decreases before large seismic events and recover its "background" level after them.

The validity of the ST as a tool to explore the correlation between tides and seismicity is subject to two basic conditions: (1) the events in the sample should be completely independent (Cochran and Vidale, 2007), i.e. a declustered sample of earthquakes should be used; and (2) the number of events should be large enough to ensure that the underlying statistical properties of the test are valid. Beeler and Lockner (2003) suggest that samples greater than ~10,000 events are required to rely on the p-value as an indicator of the null-hypothesis probability. The first condition is fulfilled by the declustering procedure applied to our full catalogue (see Section 3.2) and the statistical test we applied to the resulting dataset (see Section 3.2.2). The second condition is also fulfilled when we perform the analysis on the whole sample (~100,000 events) or at locations, such as the Bucaramanga seismic nest, where the seismicity rate is very high. However, it is much harder to achieve when dealing with the spatiotemporal battery of analysis we have designed to study the behavior of the correlation anomaly in our sample (see Section 4). In those cases, we have always computed the p-value using a "bootstrap" algorithm (the BPVC described in the next paragraph) that gives us some statistical confidence in the result, but more importantly, an estimation of the error in the p-value (see for instance the results in **Figure 11**).

In **Figure 11** we show actual visual representations of the Schuster Test for the same sample of earthquakes (intermediate depth earthquakes in central Colombia) and two different tidal wave components (monthly and fortnightly). We will call these diagrams "Schuster Walks" (this representation is like the "rose diagrams" of Heaton, 1975). We realize that for the monthly wave, a statistically significant number of events are "aligned" in the 0/360° phase indicating a correlation anomaly, at least for that component. On the other hand, the fortnightly phases of the same sample are mostly random and a correlation or the lack of it, for this periodicity, cannot be demonstrated nor discarded. The value of log $p$ in the examples of **Figure 11** is reported with an associated error. We calculate such errors using a resampling algorithm (hereafter "bootstrapping p-value calculation" or BPVC). For that purpose, we take the original subsample we are analyzing (N events) and resample it without replacement for a total of ~50 times (50 resamples of size $fN$, where $f=0.8$). The p-value is computed independently for each resampled set. The mean of log $p$ is estimated as the arithmetic average of this quantity over the resampled dataset. The estimated standard deviation (error) is the square root of its variance. Of course, in doing this we do not assume that the values of log $p$ are distributed normally. We simply use the arithmetic mean and standard-deviations as indicators of the variability of the statistics in the studied subsample.

As a reference, we will be looking for indicative p-values $p < 5\%$, or equivalently $\log p < -3$, as done by other researchers (Tanaka et al, 2002, 2006, Cadicheanu et al., 2007; Métivier et al., 2009; Vergos et al., 2015). Values closer or larger than this arbitrary threshold are not as informative of a correlation anomaly, as values significantly lower than it.

## 4. Results and discussion

To assess the existence of p-value anomalies in the sample of earthquakes in Colombia in the last two decades, we have applied a "multidimensional" strategy. Potential correlations with different wave components of the tide may arise depending on four basic factors: 1) geographical location, 2) time, 3) depth and 4) earthquake magnitude. We will explore in depth the presence (or absence) of correlation anomalies in this four-dimensional space.

In a first step, we want to study global correlation anomalies in the data. In Table 1 we present the results of calculating p-values for the whole region but different sample selections. Here, we analyze earthquakes across the study region, irrespective of their depths. Magnitude thresholds consistent with the completeness of the sample (see **Figure 6**) have been used to filter the events for this specific test.

Results shown in Table 1 are consistent among different selected subsamples. Thus, for instance, no significant correlation is suggested by the data for the semidiurnal and fortnightly components, nor for the whole period, neither in the A or B time windows. On the other hand, diurnal and monthly components suggest a correlation of p-values, at least for the range of magnitudes considered here.

| Component | Whole region, All times, M>3.0 | Whole region, Period A, M>3.0 | Whole region, Period B, M>2.0 |
|---|---|---|---|
| **Semidiurnal** | -0.9 (0.8) [5869] | -1.1 (1.0) [4465] | -1.7 (1.2) [7488] |
| **Diurnal** | -7.0 (3.3) [5787] | -5.3 (3.2) [4403] | -13.2 (4.1) [7375] |
| **Fortnightly** | -2.0 (1.8) [5689] | -1.0 (1.0) [4328] | -2.2 (1.9) [7241] |
| **Monthly** | -12.1 (4.9) [5872] | -14.4 (4.4) [4467] | -10.8 (4.4) [7494] |

*Table 1. Global p-values calculated for different subsets of our earthquake database. See Figure 3 for definition of periods A and B.*

To better understand the origin of these significant correlation anomalies and track the set of properties (depth, magnitudes, geographic regions) for which the anomalies arise, we analyze the p-value in one and two dimensional "sections" of our 4-dimensional parameter space. This means that one or two properties will be considered at a time. Thus, for instance, in **Figure 12** we show p-values for all tidal components, calculated in 25 different quadrants across the region (this is a 2-dimensional section). In all cases, we have included events at all depths and times. We have calculated p-values only for earthquakes such that $M_L>3$. **Figure 12** reveals that correlation anomalies for the diurnal and monthly components arise mainly for earthquakes around the Bucaramanga nest (lat. +7, lon. -73) and Cauca cluster (lat. +4.5, lon. -76). Those areas concentrate the events with the largest depths and magnitudes in the study area (see **Figure 2**).

In **Figures 13** and **14** we study the dependency of log $p$ on depth and magnitude (these are also two-dimensional sections of our four-dimensional property space). For that purpose, we created a grid of values ($M_L,d$). For each point in the grid we get subsamples of events fulfilling the condition depth<$d$ and magnitude<$M_L$. If the number of events in a subsample is less than 3,000, the p-value is not computed (low statistical significance). In this case, an arbitrary value log $p=0$ is assigned to the point in the grid.

Again, the semidiurnal and fortnightly components do not exhibit any significant correlation anomaly (**Figure 13**). However, in the case of the diurnal and monthly components (**Figure 14**), significantly low values of log $p$ are observed, especially for subsamples including intermediate depth events (d>150 km) and earthquakes with magnitudes larger than 3.0-3.5; the correlation anomaly seems to be significant for deep events, while shallow events show only moderately low p-values.

We also study the evolution in time of log $p$ during the last two decades in the study region (one dimensional time sections). For this purpose, we create a dozen of subsamples centered at specific dates in the time interval covered by the data. Each subsample contains the same number of events (3000); we select events having magnitudes larger than 2.0 and occurring at any depth. The results of this analysis are presented in **Figure 15**.

The evolution in time of log $p$ confirms, once more, the existence of a correlation anomaly among the diurnal and monthly components, while no significant anomalies seem to exist for the semidiurnal and fortnightly waves. More interestingly, the result in **Figure 15** reveals that the correlation anomalies identified before seems to change in time. Thus, for instance, in the case of the monthly wave, a strong anomaly is observed only in the first 3 years of data (1993-1996) and a few years later (1998-2001). In other periods, the values of log $p$, although systematically low, are marginally above the threshold log $p = -3$. The fact that the lowest values of log $p$ fall precisely in the period where the sample is more incomplete, is somewhat suspicious. To check that no "sampling" artifacts, such as the fact that the "old" network collected more data during certain periods of the month and not in others, we graphically analyzed the dispersion of the events (bottom panel in **Figure 15**). No significant sampling artifacts were identified, nor discarded. A more thorough analysis should be pursued to discard that the high correlation suggested by the data in this period is produced by other numerical or instrumental effects.

The case of the diurnal wave is interesting and may shade some light on the possibility that the correlation anomaly may be a real effect instead of a sampling artifact observed in period A. In the corresponding panel of **Figure 15**, systematically low values of log *p* are observed during a long period of time between 2003 and 2013 (which includes data from Periods A and B) with a minimum around 2007. At times earlier than 2003 and later than 2013 log *p* values are not low enough to reject the null-hypothesis of random phases. Again, it is interesting to notice that in 2007 the Colombian seismological network started the transition from older and more sparse stations to the more modern ones used in the present. The minimum in log *p* approximately coincides with this transition, suggesting a sampling artifact. Even if that is the case, the systematically low values of log *p* observed in the period 2009-2013, when the modern network was already under operation, hint to a correlation anomaly that we cannot discard.

In terms of tidal effects, this would mean that globally, in Colombia, there are ~16% more earthquakes happening close or around the time of the largest monthly tides, than in the rest of the month. Does this mean that there are more earthquakes near the perigee than during apogee (or any other time in the lunar month)? Not necessarily. Maximum tides at a given place on Earth do not necessarily coincide with the time of the perigee (as can be seen in **Figure 8**). Tide amplitudes depend, both on the position of the moon and the sun in space, as well as on the location of the earthquake and the time of day; therefore, the time during the month when larger tides occur may not coincide with the time of perigee.

It is interesting to notice that if earthquakes would happen randomly during the lunar month, the number of events occurring around the apogee should be larger. In virtue of the second Kepler's law the moon spends more time at larger distances than at shorter ones. If we consider the lunar true anomaly as the angle between the position vector of the moon and the direction of the perigee in the lunar orbit, we find that there are almost 20% more earthquakes happening when the moon is at the perigee than in the rest of the lunar month, which is also a hint of a possible earthquake-tide correlation. A plot of number of earthquakes vs. lunar true anomaly is included in the supplementary material (Figure S3).

Considering our observations and analyses, the correlation between tides and seismicity is hinted for earthquakes in northwestern South America. A definitive demonstration of such a causal relationship would only be possible by an in-depth analysis of the physics of tidal triggering, taking into account the tectonic setting of the region.

For this purpose, we could for instance recover the focal mechanism of many earthquakes, especially in the areas of interest, namely the Bucaramanga nest and the Cauca cluster, and from there we could compute the tidal Coulomb failure stress (TCFS) (see eg. Stein, 1999; Vidale et al., 1998; Cochran et al., 2004; Fischer et al., 2006; Wilcock, 2009; Xu et al., 2011; Miguelsanz del Alamo, 2016; Bucholc and Steacy, 2016). Even though the calculated focal mechanisms may not show a clear pattern in favor of certain type of failure, it would be valuable to observe the behavior of these events in relation to tides. The analysis of the TCFS for a significant number of earthquakes, and more importantly, the correlation between this value and the phases as calculated in this work, could provide a clearer indication of tidal triggering in north-western South America. For the sake of brevity, however, we have decided to leave this in-depth analysis for a future paper.

Most of our conclusions are based on the observation of low p-values in the Schuster test. However, it is important to stress again that a small p-value is not a definitive indication of a correlation between tides and seismicity. The reason of this general misconception arises from an incorrect interpretation of probabilities in this and in similar cases. The probability that earthquakes be triggered by tides (lets abbreviate it as TT) given the observed events (E), symbolically $P(TT|E)$, is not the same as the probability that the observed events be drawn from a random instead of a periodic process, $1-P(E|TT)$. The latter is precisely what we called the p-value. For a recent discussion on the actual interpretation of the p-value see Wasserstein and Lazar (2016) and references there in.

Some of the results obtained by applying the ST to heterogeneous samples of earthquakes, i.e. earthquakes collected and analyzed by a network whose quality and distribution change in time, may favor the existence of artificial correlations. Thus, for instance, if the sampling or the analysis of the network has any kind of periodicity, these effects may produce numerical artifacts in the resulting phases. A proper analysis of these effects must be pursued. We have attempted to check as carefully as possible that none of our results arose from those artifacts. An important test to check if the correlation anomalies reported before could arise from artifacts in our analytical tools was done by creating mock samples of earthquakes having, on one hand, the same location of the true ones but a random different time and date (*mock-time sample*) and, on the other hand, the same date and time but a random location around the actual hypocenter of the real events (*mock-space sample*). For the mock samples we conducted a similar analysis as that presented in Figures 13 and 14; in contrast to the actual database, the mock sample does not show any significant correlation anomaly, demonstrating that the anomalously low values of log $p$ observed in **Figure 14** are a true property of the sample and not a numerical artifact arising by the application of our methods, and that random earthquakes analyzed with our methods and tools do not exhibit any phase clustering as strong as the actual earthquakes in our database. Results of the analysis for the *mock-time sample* are shown in supplementary Figure S4.

Finally, the Moon may provide some clues for solving the puzzle of tidal-triggering on Earth. Recent reanalysis of the data collected by four seismometers installed by the Apollo Mission between 1969 and 1977, revealed a significant correlation between deep moonquakes and terrestrial-solar tides (Nakamura, 2005; Bulow, 2006, 2007; Weber, 2009). If deep moonquakes are triggered by earth tides, there is no reason to think that intermediate to deep earthquakes may not also be triggered by lunisolar tides.

# 5. Towards an earthquake and tides information system

Analyzing the correlation between tides and seismicity is not a trivial numerical task. Not only the software required to compute synthetic tides such as ETERNA are robust and very complex, but the amount of computational resources involved at storing and analyzing tidal time series for hundreds of thousands of events is significant. Reproducibility is also becoming a matter of concern in many fields and this should not be the exception.

We propose the development of a computational platform (information system) able to fulfill some of the needs of this active area of research. The information system should be designed for providing access to the results of simulations to a wider audience and to researchers in other fields. The key objectives of such an information system could be:

1) To provide access to all the relevant information available about the earthquakes in the database (time, location, magnitude, focal mechanism, associated errors, etc.).
2) To provide access to the tidal time series around the time of each earthquake in the database.
3) To include basic analytical tools providing the user with the capability to perform statistical analysis, basic calculations and plots with the stored information.
4) To be flexible enough to allow that custom databases can be uploaded and analyzed.
5) To be publicly available (preferably open source).

With this in mind, we developed and used for the purpose of this work, the computational tool **tQuakes**, an information system intended for storing, analyzing and making public information about the connection between tides and seismicity.

The development of **tQuakes** was simultaneous with the scientific development of this work and allowed our research team to collaborate more efficiently, while sharing the results in a friendly platform. It also allowed us to store partial results and see the evolution of the work in a way that was instrumental for understanding key aspects of the phase calculations. Most of the plots included in this work are by-products of the analytical tools of **tQuakes** and may be reproduced by users of the information system.

Although **tQuakes** is at an early development stage (only the alpha version has been released), the basic capabilities of the system (browsing the list of earthquakes and checking the results of the analysis over them) are now accessible. All the geophysical community is welcome to contribute with its development. We will be upgrading the results presented here and publishing similar results for other regions in South America in the forecoming future.

For the sake of brevity, a detailed description of **tQuakes** and many of its applications is left for a future paper. In the meantime, a development version of the information system is freely accessible from http://seap-udea.org/tQuakes while the full source code is available at http://github.com/seap-udea/tQuakes.

## 6. Summary and conclusions

In this work, we presented the first systematic exploration of the potential correlation between tides and seismicity in northwestern South America, focusing in Colombia. For this purpose, we compiled a large set of earthquakes provided by the RSNC, most of them coming from the Bucaramanga nest and the Cauca cluster, in the time interval 1993-2017. We performed basic statistical analyses over the whole database, including a declustering procedure using the Reasenberg algorithm. Our analysis showed that our database is complete down to magnitude 3.0 and follows a Gutenberg-Richter law with b=0.94, which is consistent with the complex tectonic environment in Colombia.

For each earthquake in our database we compute time series of most of the components of the tides (vertical and horizontal displacement and strain, tidal acceleration, tilt, etc.) in a period of 90 days before and after the time of the events. We devised novel methods to compute from those time series the phase of the tidal components (semidiurnal, diurnal, fortnightly and monthly) at the earthquake origin time. Our methods (especially in the case of the diurnal and monthly phases) rely on astronomical information to find the proper reference time with respect to which we compute the corresponding phase. We have verified that our approach, in contrast to other methods based only in the signal itself, creates a stronger correlation between the value of the phase and the actual state of tidal stress.

Despite some negative preliminary results available in the literature, we found a significant correlation anomaly with the diurnal and monthly phases, especially for intermediate depth earthquakes in the Bucaramanga nest and the Cauca cluster. The anomaly suggests that on average 16% more earthquakes are triggered when the monthly phase is between $350^o$-$10^o$, which corresponds to the time in the lunar month when the moon is closer to the Earth (perigee) and tides have larger maxima. These results are consistent with the findings of studying deep moonquakes that are also triggered preferentially during the closest approach between the moon and the earth.

Although we did not perform exhaustive statistical tests to discard that the correlation anomaly may arise from sampling artifacts, for instance, from the fact that the quality of our database is somewhat heterogeneous (the infrastructure of RSNC changed around 2007), we tested that our results are not a product of the way we analyzed the tidal time series.

All our analysis and calculations have been performed using **tQuakes**, a new information system designed and developed specifically for this work that compiles all the relevant information about earthquakes in a target database, including information about the tides around the time and place of each earthquake. The information system is publicly available over the web in its alpha version, and its source code is open. **tQuakes** is the seed of an effort to create tools for making this area of research more accessible to a wider research audience but also to increase the level of reproducibility in the field, which is a growing concern in many areas of research.

**Acknowledgements**

Most of the computations that made possible this work were performed with Python and its related tools and libraries, Matplotlib (Hunter et al. 2007), scipy and numpy (Walt et al. 2011). We thank the Geological Survey of Colombia (SGC for its acronym in Spanish) for the freely available dataset of earthquakes. Tides were calculated with ETERNA and we appreciate the support and availability of Thomas Jahr and Adelheid Weise at the Jena University, who helped us with the usage of the package, but also helped us to fully understand the limitations of the software. This work is supported by Vicerrectoria de Docencia-UdeA and the *Estrategia de Sostenibilidad de la Universidad de Antioquia*. G. Moncayo is supported by Colciencias, Doctorado Nacional - 727 program. G. Moncayo thanks the University Friederich-Schiller in Jena for the hospitality during a mid-term visit when a significant part of this work was developed. We are especially thankful to Diego Olmos, whose insightful questions during the lectures of planetary sciences and geophysics, originally motivated us to pursue these questions.

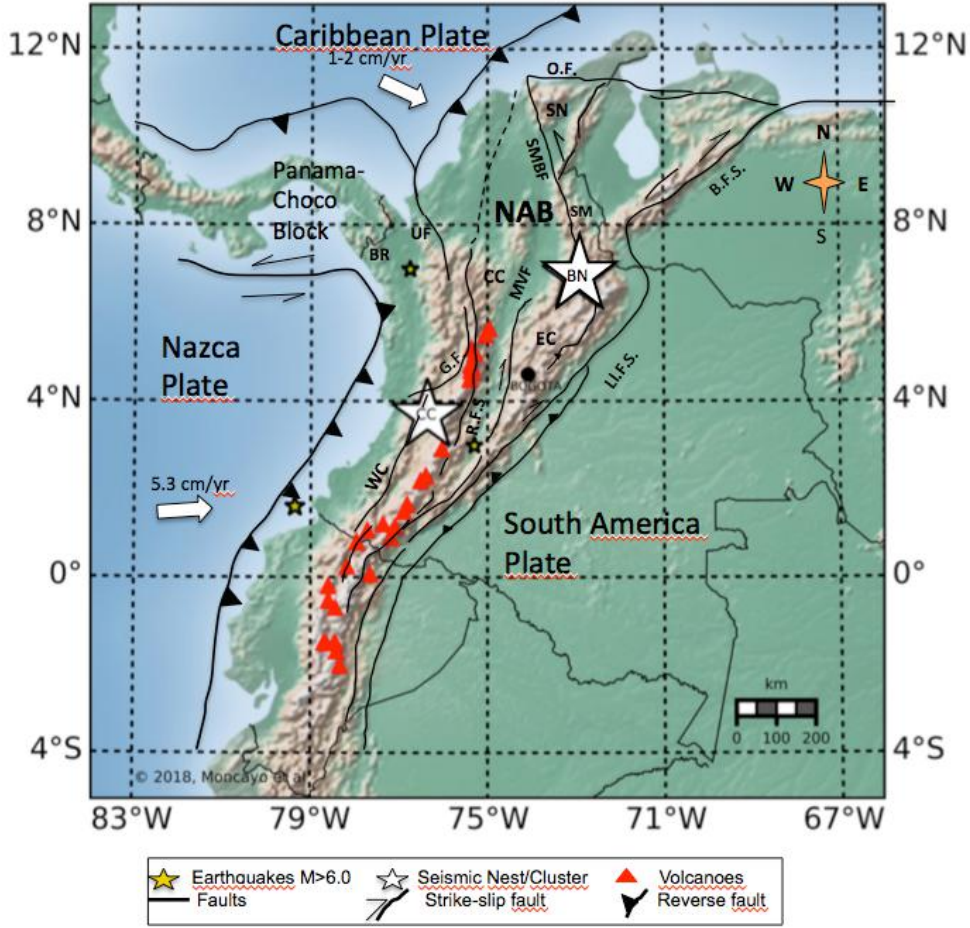

*Figure 1. Main tectonic features of Northwestern South America (after Trenkamp et al., 2002; Pulido et al., 2003; Cortes and Angelier, 2005). NAB: North Andean Block; WC: Western Cordillera; CC: Central Cordillera; EC: Eastern Cordillera; CC: Cauca Cluster, NB: Bucaramanga Nest; R.F.S.: Romeral Fault System; Ll.F.S.: Llanos Fault System; S.M.B.F.: Santa Marta-Bucaramanga Fault; G.F.: Garrapata Fault; UF: Uramita Fault; BR: Baudó Range; MVF: Magdalena Valle Fault System. O.F.: Oca Fault; B.F.S.: Boconó Fault System. Red triangles represent active volcanoes.*

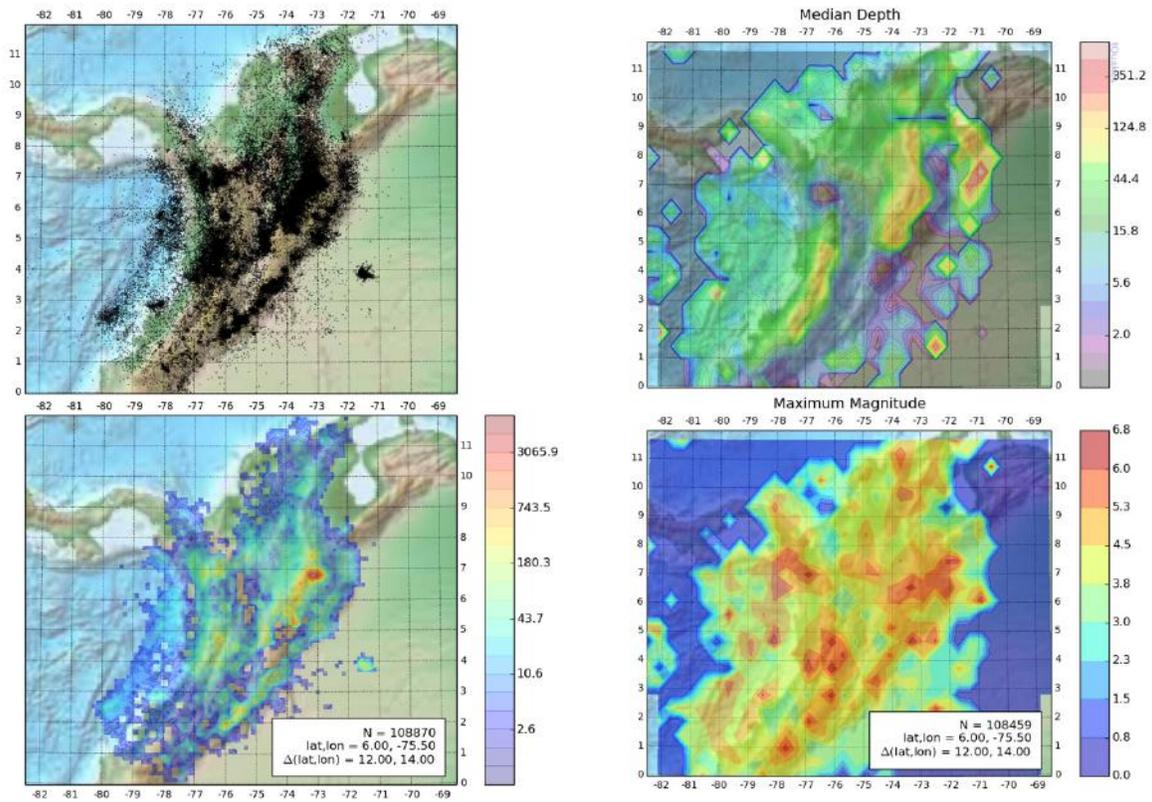

*Figure 2. Maps of distribution and properties of declustered earthquakes in Colombia between 6 June 1993 and 13 August 2017. Top left: Earthquake epicenters. Bottom left: geographic density of earthquakes. Top right: Median of earthquake depth. Bottom right: Maximum recorded local magnitude. Note the cluster of earthquakes at latitude ~4°, and longitudes between -71° and -72°; this cluster was not known before 2013.*

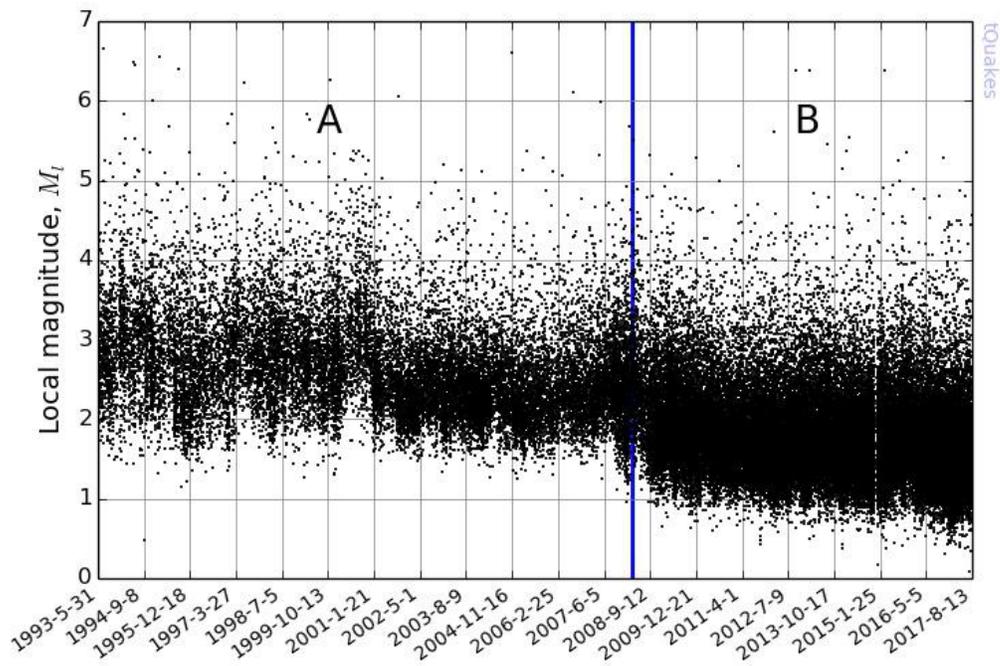

*Figure 3. Scatter plot of the time and local magnitude of the declustered sample of earthquakes considered in this paper. Periods A and B are defined for the times before and after the densification of the seismological network.*

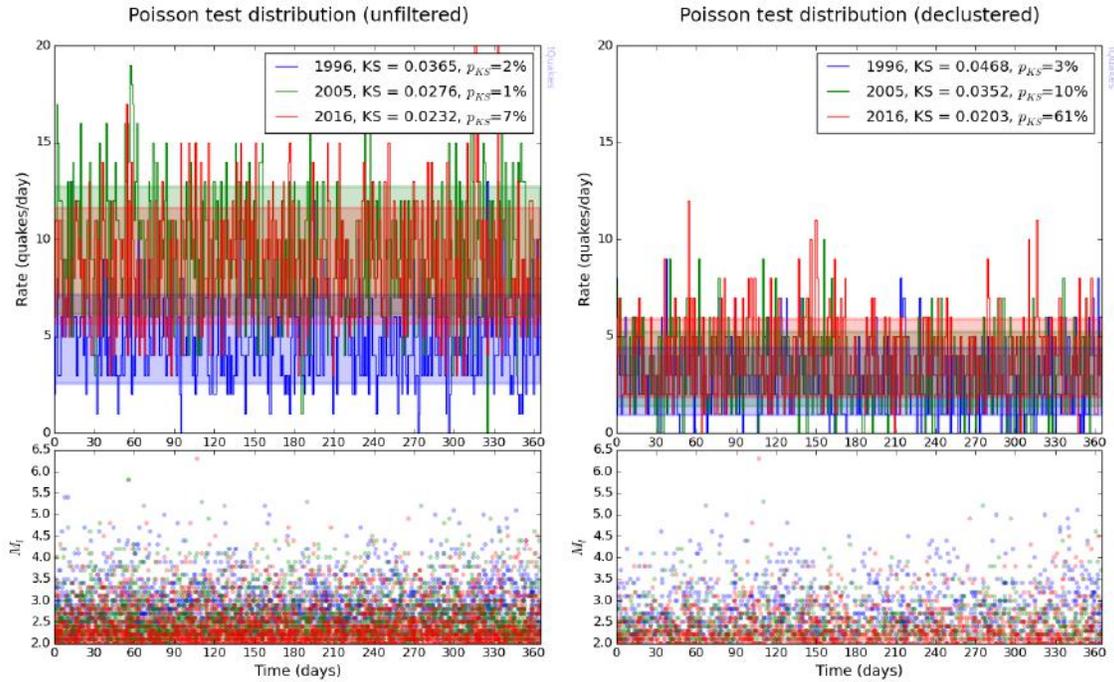

*Figure 4. Poisson test for selected samples in the case of the entire catalogue (left panels) and the declustered dataset (right panels). For this test, we took three years of earthquake data (1996, 2005 and 2016), each of them denoted by a distinct color. For each year, we calculated the Kolmogorov-Smirnov statistic (KS) and its corresponding p-value ($p_{ks}$).*

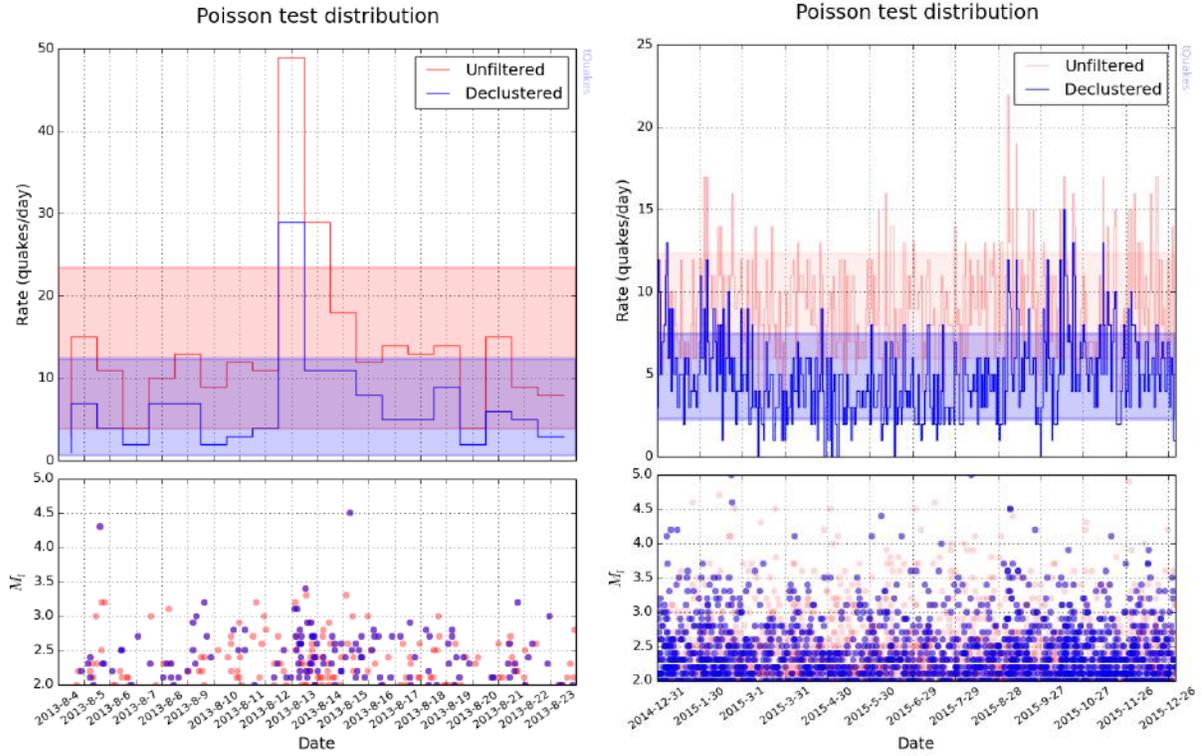

*Figure 5. Poisson test for two patologic cases: left panel represents the case when the declustering algorithm does not remove the foreshocks: see the event with magnitude Ml=4.5 (bottom left) right after the peak in the rate of seismicity (top left); the right panels illustrate those cases where the declustering algorithm reduces the number of events below the average rate ("swinging" behavior of the declustered event rate in the first half of the window).*

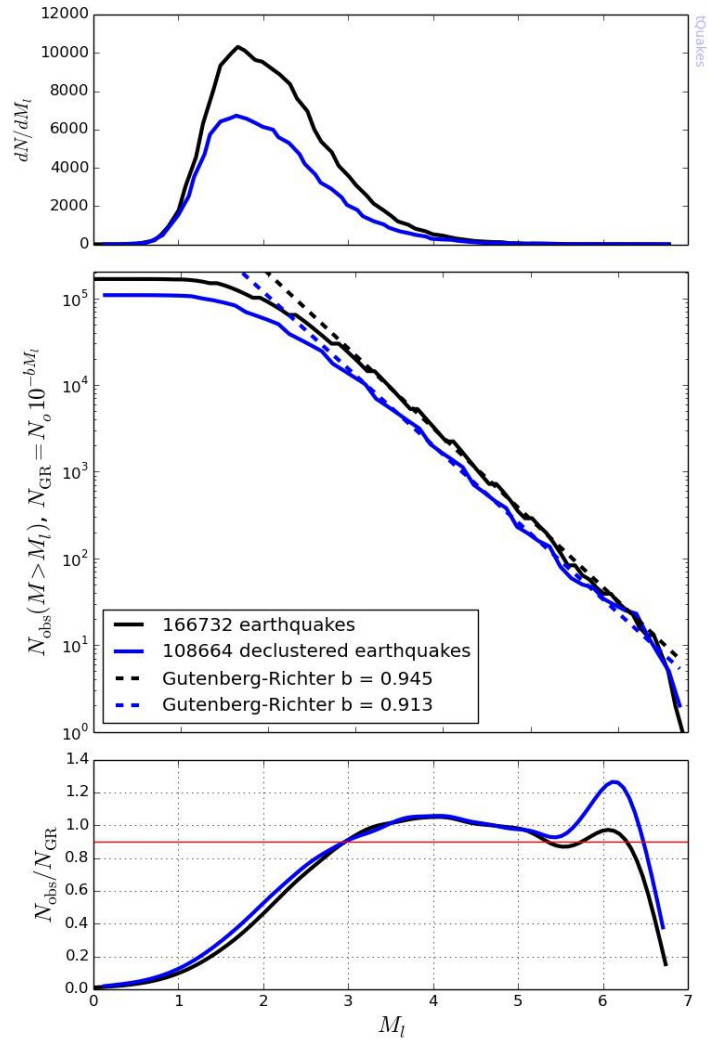

*Figure 6. Earthquake local magnitude distribution. Top panel: probability density distribution. Middle panel: cumulative distribution of earthquakes (b-value plot), including the best fit straight line that follows the Gutenberg-Richter Law (dashed curves). Bottom panel: ratio between the observed number of earthquakes below a certain magnitude, and the number or earthquakes predicted by the best fit Gutenberg-Richter Law. Black curves correspond to the entire event database while blue lines correspond to the declustered sample of the dataset.*

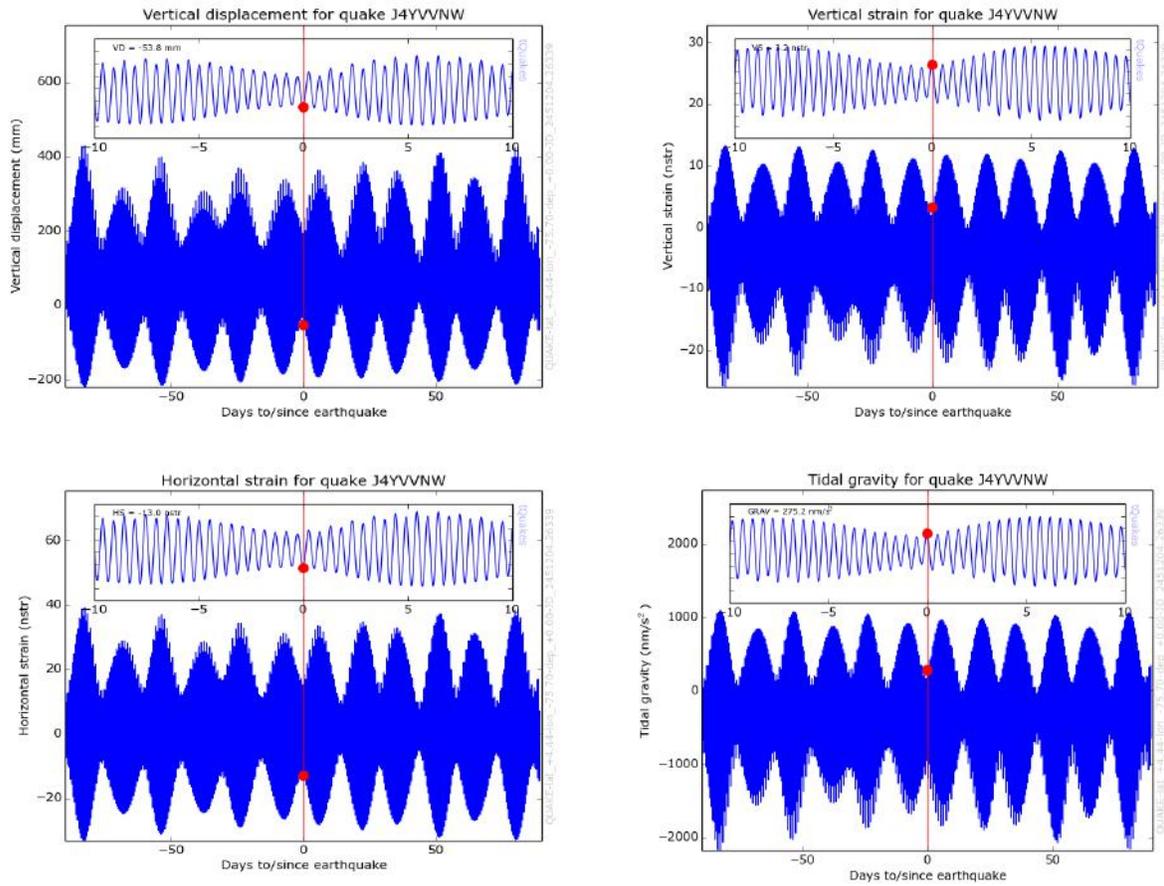

*Figure 7. Examples of the time series for the synthetic tide for the large and shallow event ($M_L$ = 6.3, depth<14.8 km) that affected people and properties in the 1999 Quindio Event. For illustration purposes, we show only the time series for four of the components of the tide: vertical displacement in mm (top left), vertical strain in multiples of $10^{-9}$ (nanostrain, top right), horizontal strain in the direction of Azimuth 0 (bottom left) and tidal acceleration in nm/s$^2$ (bottom right). The inset panel shows the tidal signal 10 days before and after the event. The red dot shows the value of the signal at the time of the earthquake.*

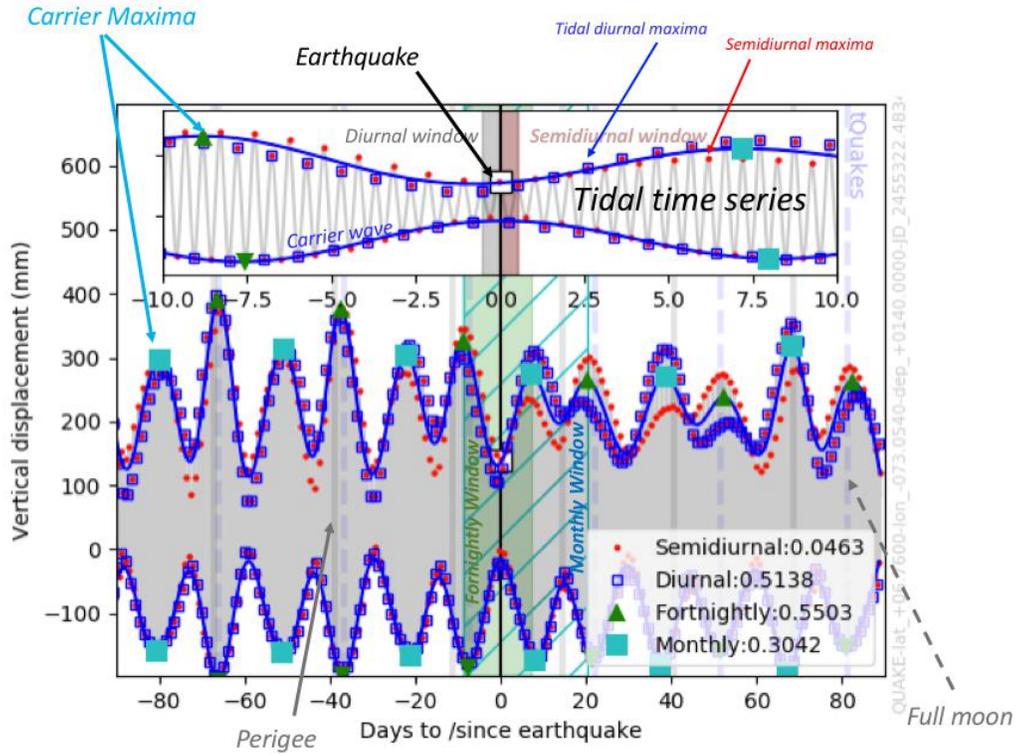

*Figure 8. Illustration of the novel method used in this work to compute the phases of the principal components from the time series of the tide around the time of an earthquake (see text for a detailed explanation). The inset panel shows the tidal signal 10 days before and after the event. The values of the tidal phases (legend panel) are dimensionless in the interval [0,1] (see text for an explanation).*

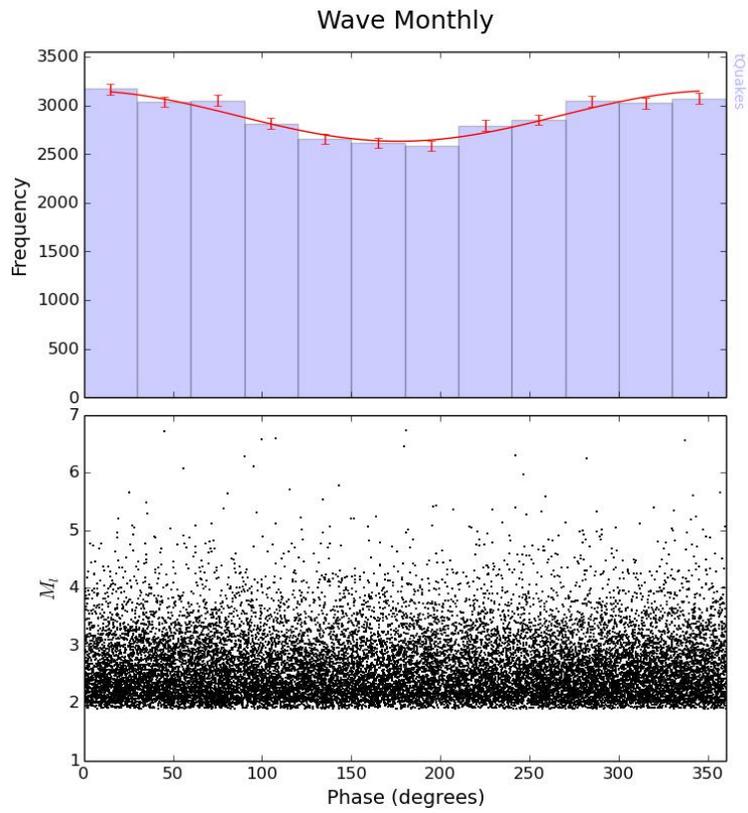

*Figure 9. Distribution of phases calculated for the monthly wave of 34,706 earthquakes in the 1993-2015 period in the study area (all depths and magnitudes are larger than 2.0).*

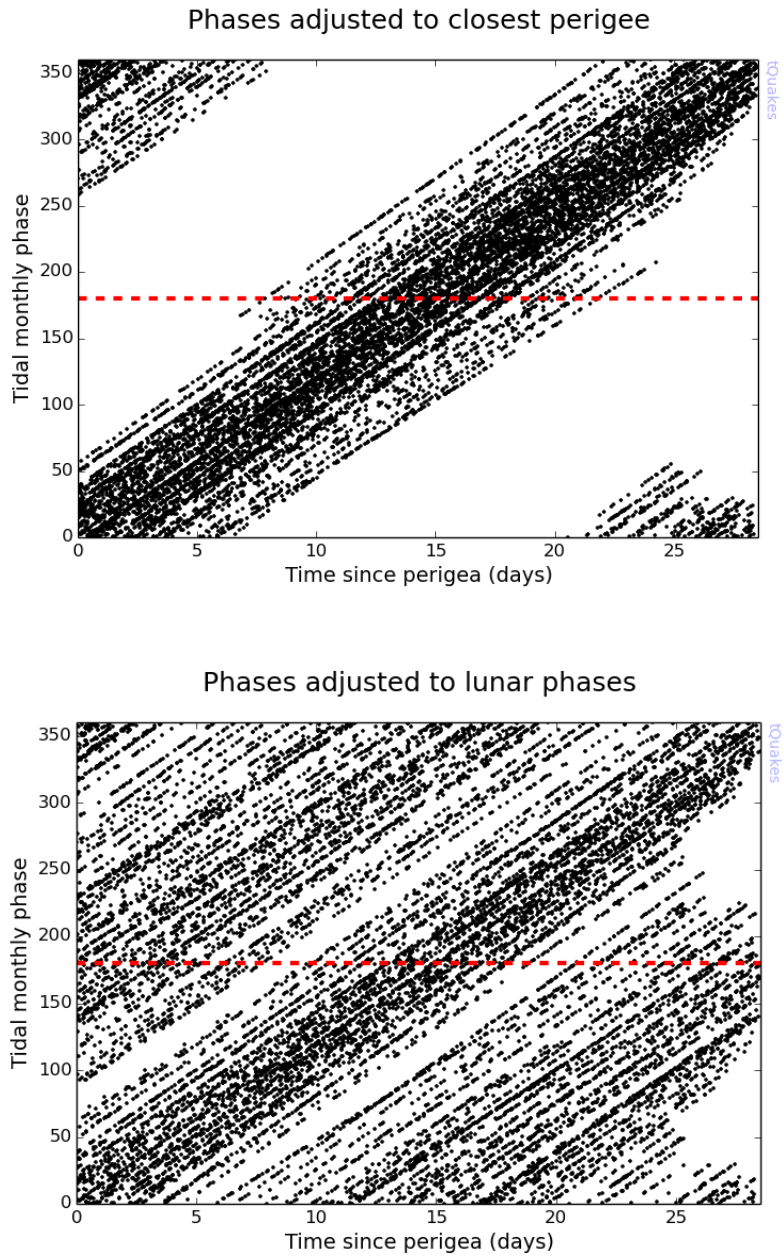

*Figure 10. Correlation between the monthly phase and the time since perigee in a subsample of events in our database. Upper panel: monthly phases calculated using a definition of the phase consistent with what has been widely used in the literature, namely, calculated with respect to an arbitrary event during the lunar month (in this case with respect to the maximum of the carrier wave closest to the last full moon). Lower panel: monthly phases calculated with our method. The red dashed line shows the phase that corresponds to the middle of the period.*

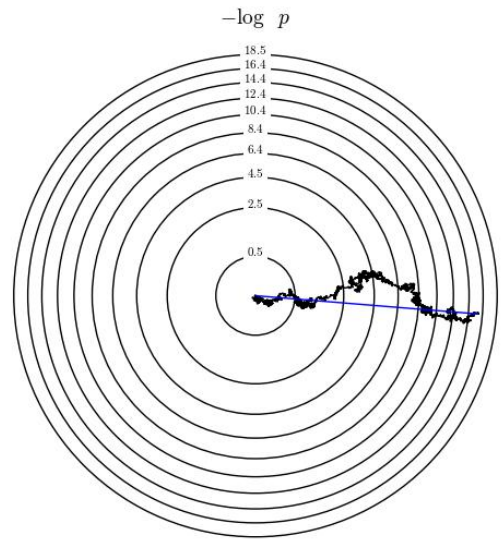

$N=4645$, log $p=-13.7\pm4.2$, Vertical displacement, Wave Monthly

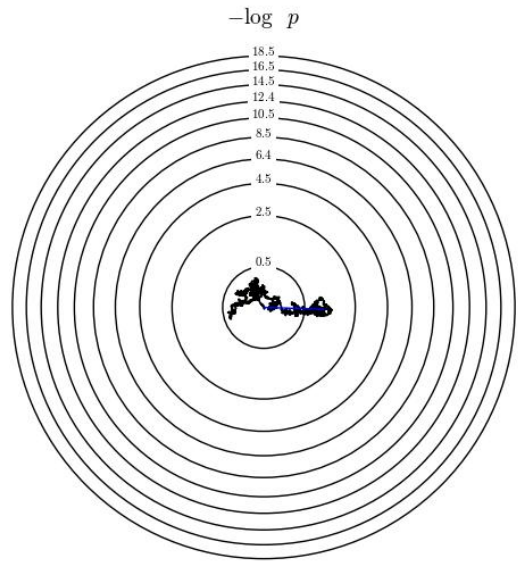

$N=4491$, log $p=-1.6\pm1.3$, Vertical displacement, Wave Fornightly

*Figure 11. Schuster Walks (see text) for a sample of intermediate depth earthquakes in central Colombia. Upper panel: Schuster walk of the monthly phase. Lower panel: Schuster walk of the fortnightly phase. The value of log p and its error are calculated by taking the whole earthquake sample.*

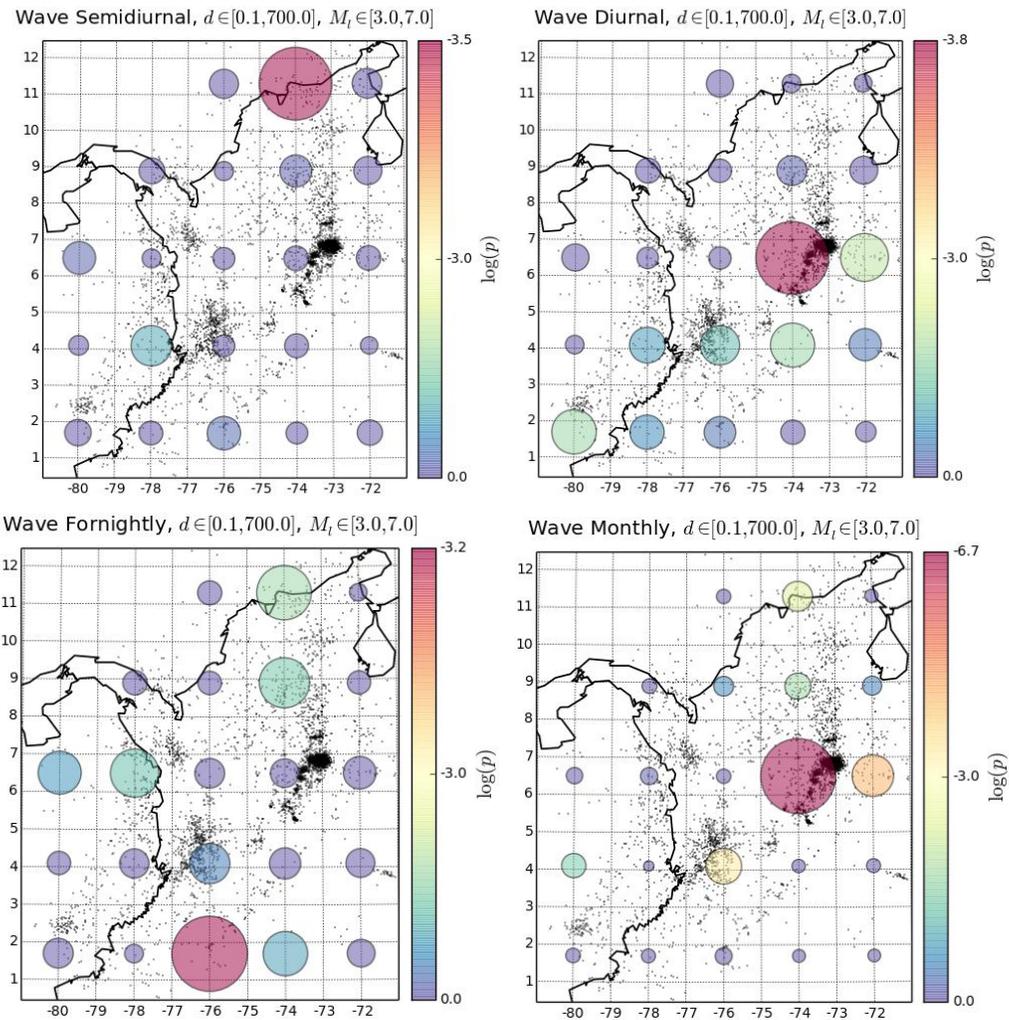

*Figure 12. Geographical distribution of the p-values as computed in 25 quadrants distributed across the study region. The p-value is coded using a diverging color map in such a way that the critical value log(p)=-3 correspond to light yellow. The sizes of the circles at the center of each quadrant are proportional to the absolute value of log(p). Top left: Semidiurnal component of the tide. Bottom Left: Fortnightly component. Top right: Diurnal component; Bottom right: Monthly component.*

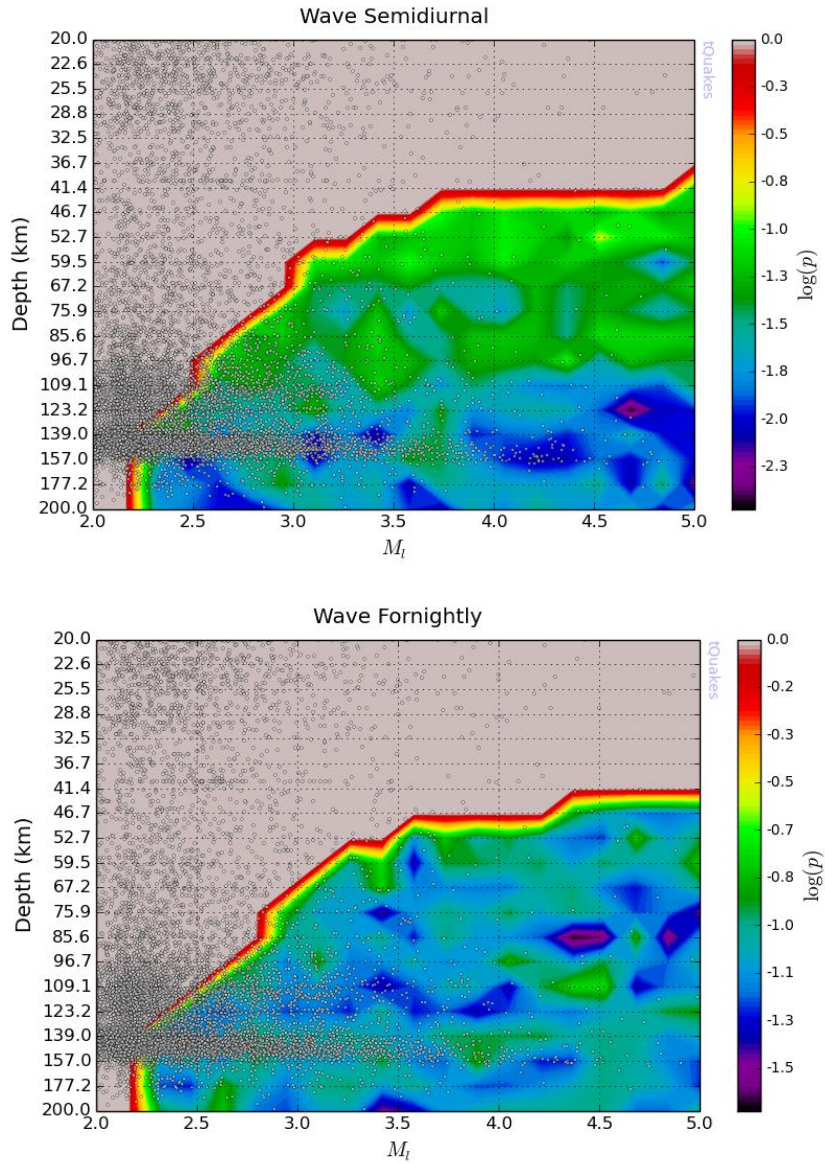

*Figure 13. Dependency of* log *p with depth and earthquake local magnitude in the case of the semidiurnal (top panel) and fortnightly components (bottom panel). Each color in the ($M_L$,d) plane, represents the value of* log *p calculated for all the events such that depth is less than d and magnitude is less than $M_L$. The gray regions correspond to values of $M_L$ and d such that the corresponding subsample has less than 3000 events and the value of* log *p is less statistically significant. The small white circles represent the depth and magnitude of each event in the full sample. The events used here are those for which the error in depth is lower than the absolute value of depth.*

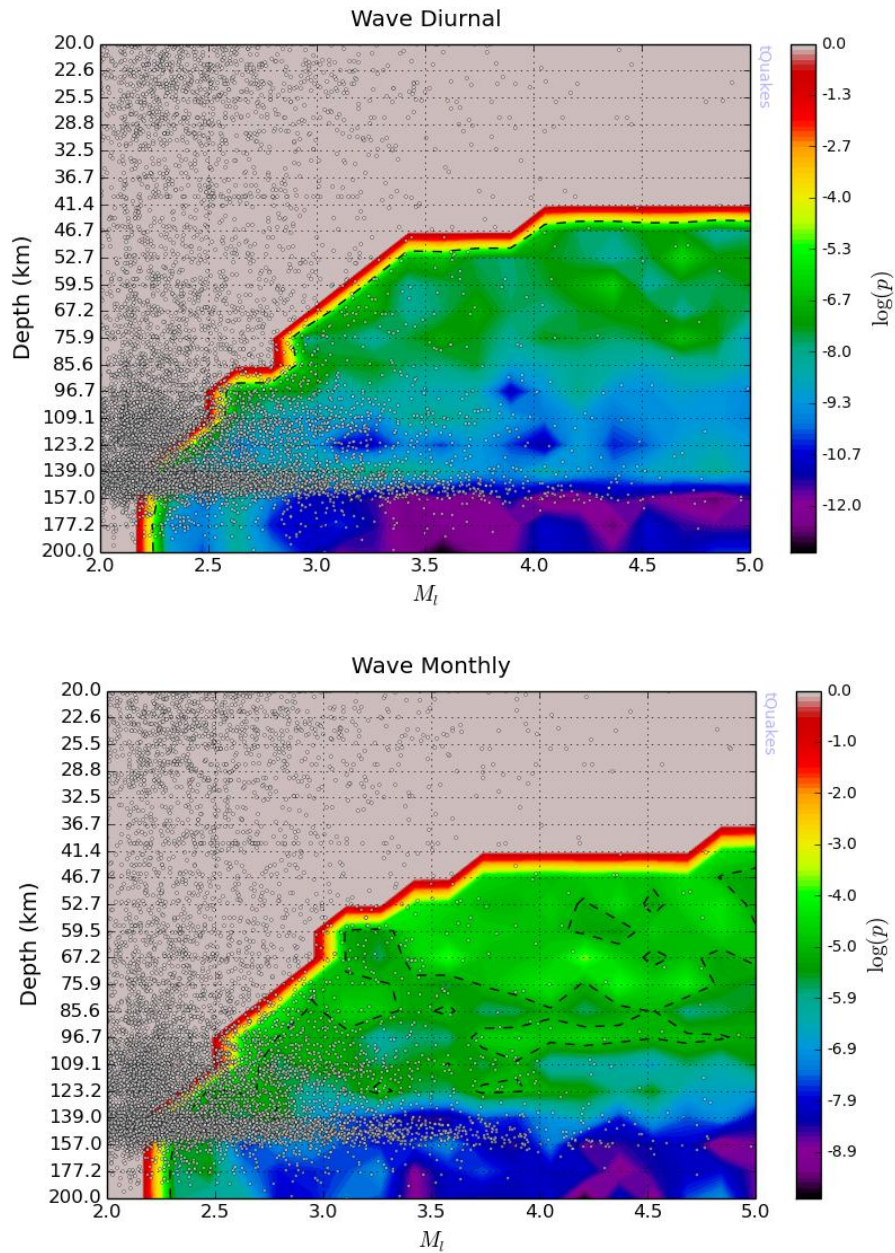

*Figure 14. Same as Figure 13 but for the diurnal (top panel) and monthly components (bottom panel). Dashed lines indicate points in the grid where the critical value* log *p=-3 is achieved.*

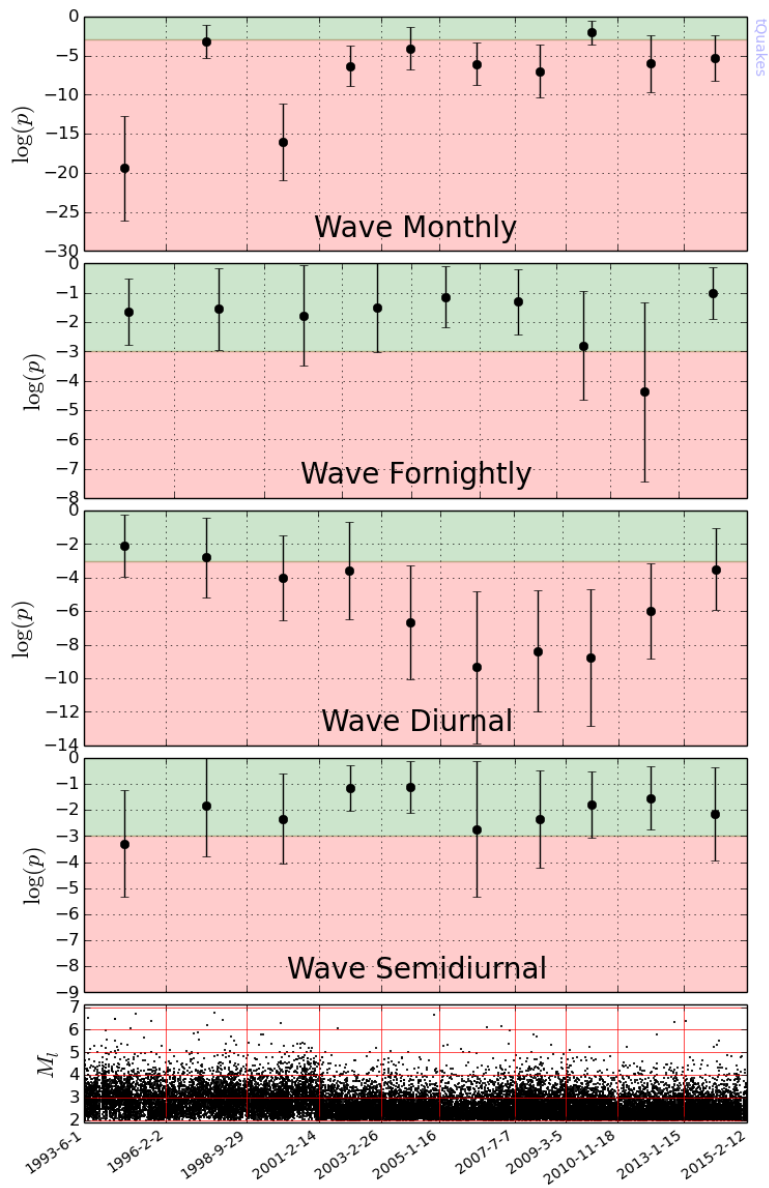

*Figure 15. Evolution of* log *p between 1993 and 2015 in Colombia. Each dot corresponds to the value of* log *p for a subsample of 3,000 events around a given central date. Earthquakes were selected such that $M_L>2.0$. Errors were computed using the resampling procedure described in the text (BPVC). Areas where log p is greater or less than -3 are colored differently.*

**Supplementary Material**

In the following we provide supplementary figures to the main text.

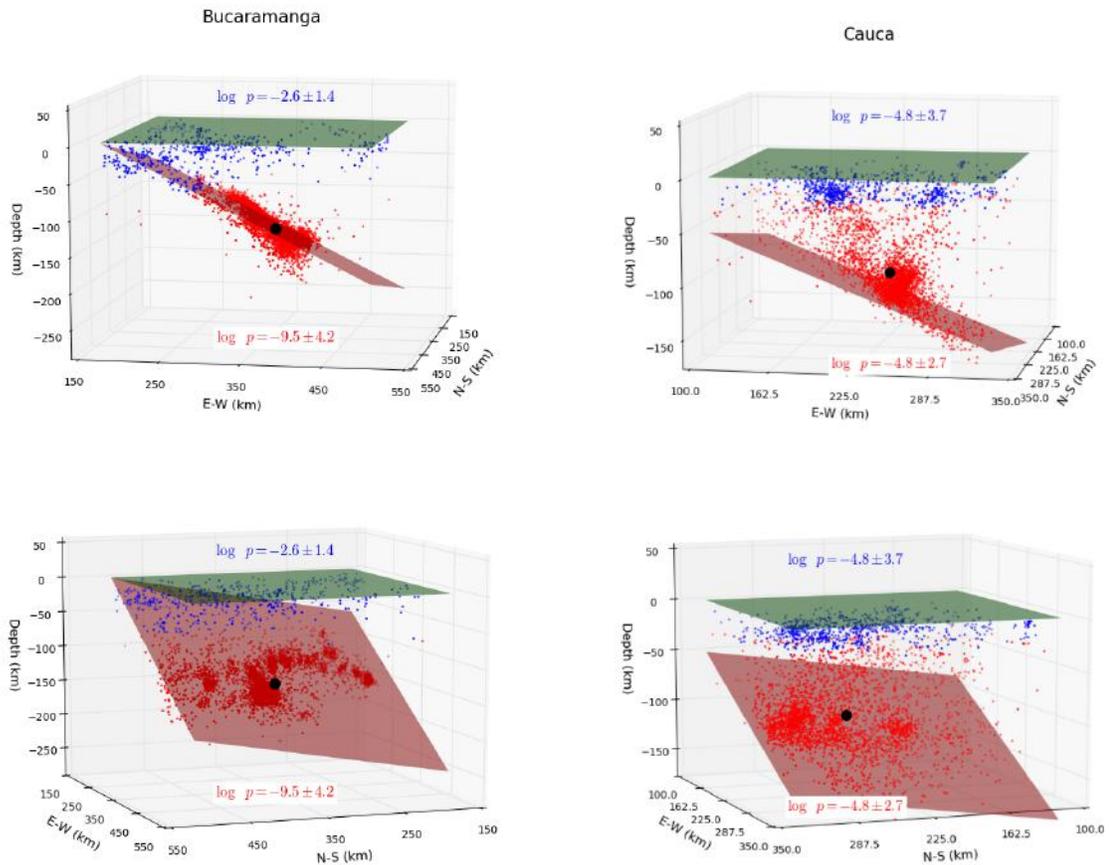

*Figure S1. Spatial distribution of earthquakes around the Bucaramanga seismic nest (lat. 6.8 N, lon. 73.1 W) and Cauca cluster (lat. 3.5 N, lon. 76.5 W). Red dots indicate the position of intermediate-depth events (depth>70 km) while blue dots correspond to shallow events. The green horizontal plane represents Earth's surface, while the red tilted plane is obtained by a least-square fitting to events with depth > 70 km. The black dot indicates the 'barycenter' of those events. The Schuster-test p-value for deep and surficial events is shown for shallow earthquakes (above the surface, in blue) and for earthquakes of intermediate depth (below the red plane, in read text).*

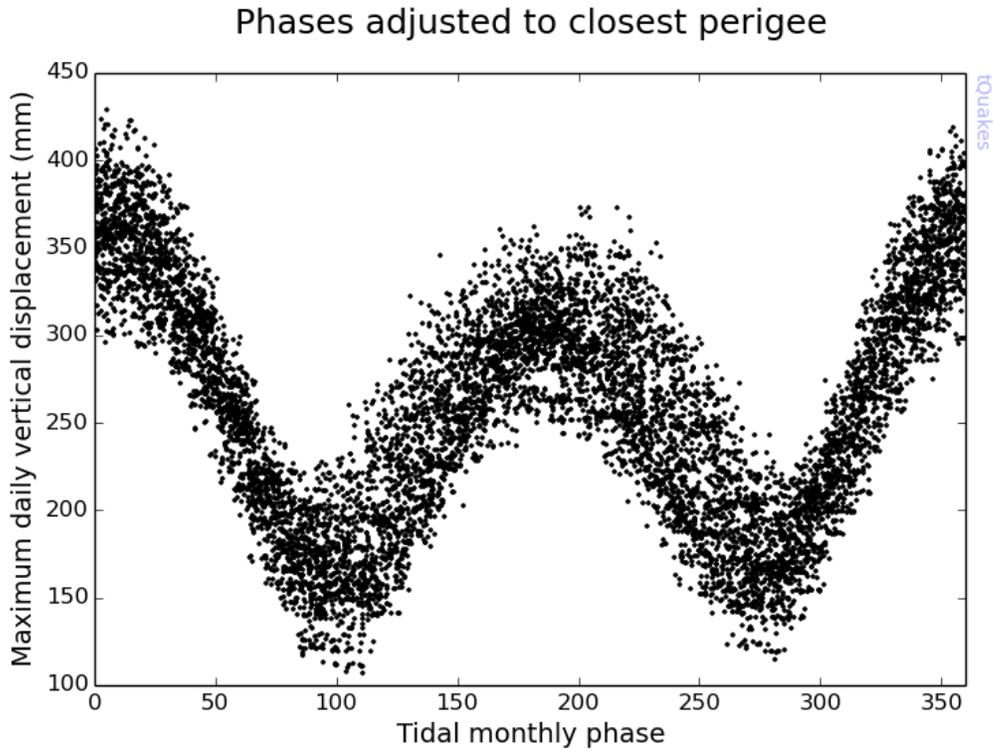

*Figure S2. Maximum daily vertical displacement around the origin time of earthquakes in our sample as a function of the monthly phase, as calculated with our method. As expected, those earthquakes having the lowest phase value (around 0 and 360 degrees) occur around days when the vertical displacement is up-to 20% larger than during the rest of the lunar month.*

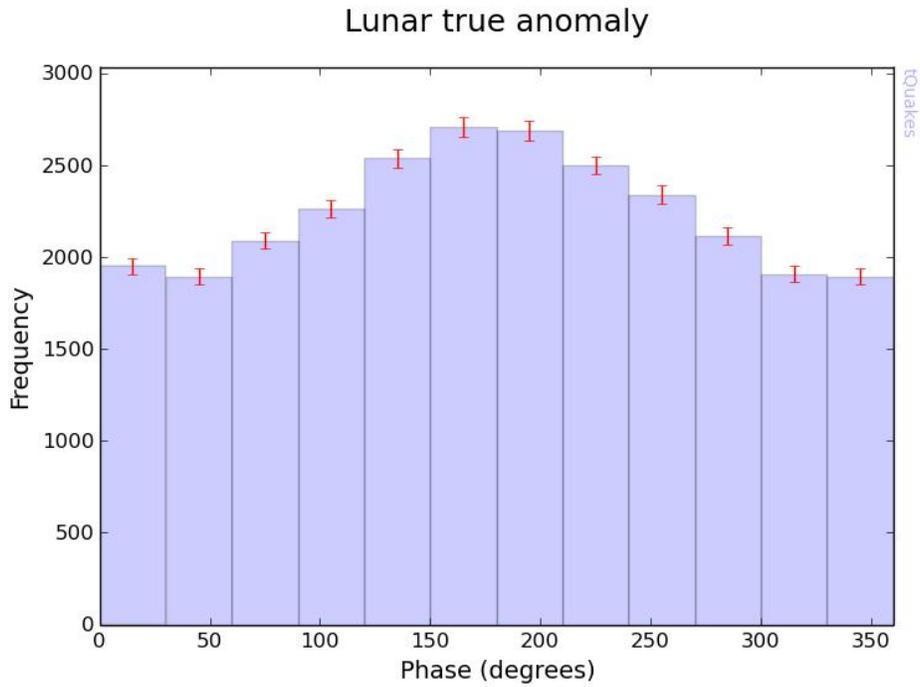

*Figure S3. Histogram of lunar true anomaly for 26,904 events in the 1993-2015 period in Colombia around the Bucaramanga seismic nest (magnitudes larger than 2.0). True anomalies close to 180 degrees correspond to positions in the lunar orbit around the apogee, where the moon spends more time in virtue of the Kepler second's law.*

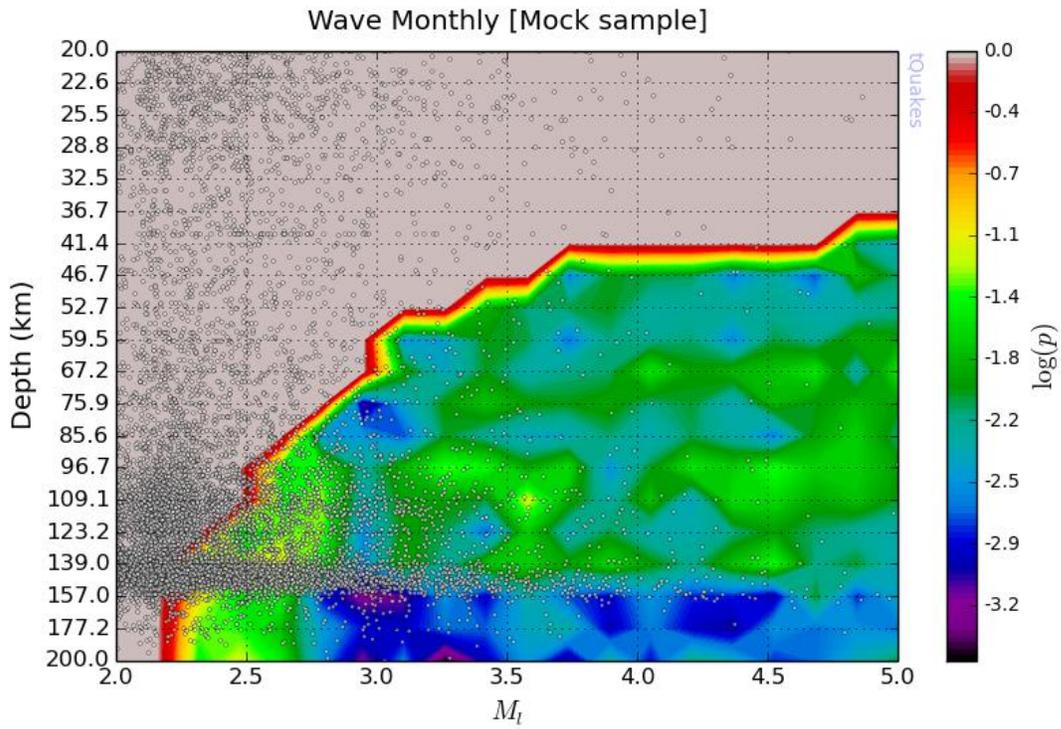

*Figure S4. Dependency of log p with depth and local magnitude in the case of the monthly component (as in Figures 13 and 14 in the main text) but for a mock sample of earthquakes for which magnitude and location were the same as in the original database but the date and time were randomly generated with a normal distribution around the actual date and time of the event (standard deviation of 10 days).*